\documentclass[a4paper,10pt]{llncs}
\usepackage{times}
\usepackage{amsfonts}
\usepackage{amsmath}
\usepackage{amssymb}
\usepackage{mathrsfs}
\usepackage{graphics}
\usepackage{graphicx}
\usepackage[utf8]{inputenc}
\usepackage{fancyvrb}
\usepackage{listings}
\usepackage{color}
\usepackage{paralist}
\usepackage{xspace}
\usepackage{url}
\usepackage[T1]{fontenc}
\usepackage[bookmarks=false]{hyperref}
\usepackage[arrow, matrix, curve]{xy}

\newcommand\fheader[1]{\hfill{\scriptsize Feature \sc #1\vskip -1.5ex}}
\newcommand\fieldsep{\circ}

\definecolor{new}{rgb}{0.9,0.6,0.8}
\definecolor{changed}{rgb}{0.9,0.8,0.2}

\def\ffj{FFJ\xspace}

\def\ffjpl{FFJ$_{\mathit{PL}}$\xspace}
\def\fj{FJ\xspace}

\spnewtheorem{thm}{Theorem}[section]{\sc}{\rm}
\spnewtheorem{lem}{Lemma}[section]{\sc}{\rm}
\spnewtheorem*{case_}{Case}{\sc}{\rm}
\spnewtheorem*{subcase}{Subcase}{\sc}{\rm}

\newcommand\f{\mathcal{F}}
\newcommand\g{\mathcal{G}}

\newcommand\qtm[2]{\man{#1}\masf{.#2}}
\newcommand\qt[2]{$\man{#1}\masf{.#2}$}

\newcommand\code[1]{{\sf #1}}
\newcommand\feature[1]{{\sc #1}}

\newcommand\masf[1]{\mbox{\textsf{#1}}}
\newcommand\man[1]{\mathrm{#1}}
\newcommand\mait[1]{\mbox{\textit{#1}}}
\newcommand\mol[1]{\overline{\masf{#1}}}
\newcommand\ol[1]{$\overline{\masf{#1}}$}

\newcommand\fieldsX[2]{\mait{fields}(\qtm{#2}{#1})}
\newcommand\fieldsY[1]{\mait{fields}(#1)}

\newcommand\mtypeX[3]{\mait{mtype}(\masf{#1}, \qtm{#3}{#2})}
\newcommand\mtypeY[2]{\mait{mtype}(\masf{#1}, #2)}
\newcommand\mbodyA[3]{\mait{mbody}(\masf{#1}, \qtm{#3}{#2})}
\newcommand\mbodyB[2]{\mait{mbody}(\masf{#1}, #2)}

\newcommand\introduceX[3]{\mait{introduce}(\masf{#1}, \qtm{#3}{#2})}
\newcommand\introduceY[2]{\mait{introduce}(\masf{#1}, #2)}
\newcommand\overrideX[4]{\mait{override}(\masf{#1}, \qtm{#3}{#2}, #4)}
\newcommand\overrideY[3]{\mait{override}(\masf{#1}, #2, #3)}
\newcommand\pred[2]{\normalfont\mait{pred}(\qtm{#2}{#1})}
\newcommand\last[1]{\normalfont\mait{last}(\masf{#1})}

\newcommand\fct{\!\rightarrow\!}
\newcommand\xss{\left[\overline{\masf{x} \mapsto \masf{s}} \right]}
\newcommand\m[1]{$\masf{#1}$}

\newcommand\always{always}
\newcommand\validref{validref}

\lstdefinelanguage{ar}[ANSI]{c++}%
  {morekeywords={refines,super,Super,original,this,layer,pointcut,call,aspect,execution,advice,around,before,after,execution,this,target,within,args,declare,parents,implements,throws,returning,throwing,throw,synchronized,boolean,proceed,cflow,cflowbelow,null,abstract,extends,interface,instanceof,privileged,cclass,document,grammar,overrides},%
   sensitive=f
   }[keywords,comments,strings]%

\begin{document}

\begin{center}
~\\[1cm]
\huge{Type-Safe Feature-Oriented\\Product Lines}
\\[2cm]
\large{Sven Apel$^\dagger$, Christian K{\"a}stner$^\ddagger$, Armin Gr{\"o}{\ss}linger$^\dagger$, and\\Christian Lengauer$^\dagger$}
\\[1cm]
\normalsize
$^\dagger$ Department of Informatics and Mathematics, University of Passau\\\texttt{$\{$apel,groesslinger,lengauer$\}$@uni-passau.de}\\[0.2cm]
$^\ddagger$ School of Computer Science, University of Magdeburg\\\email{kaestner@iti.cs.uni-magdeburg.de}\\[3cm]

\includegraphics[scale=0.2]{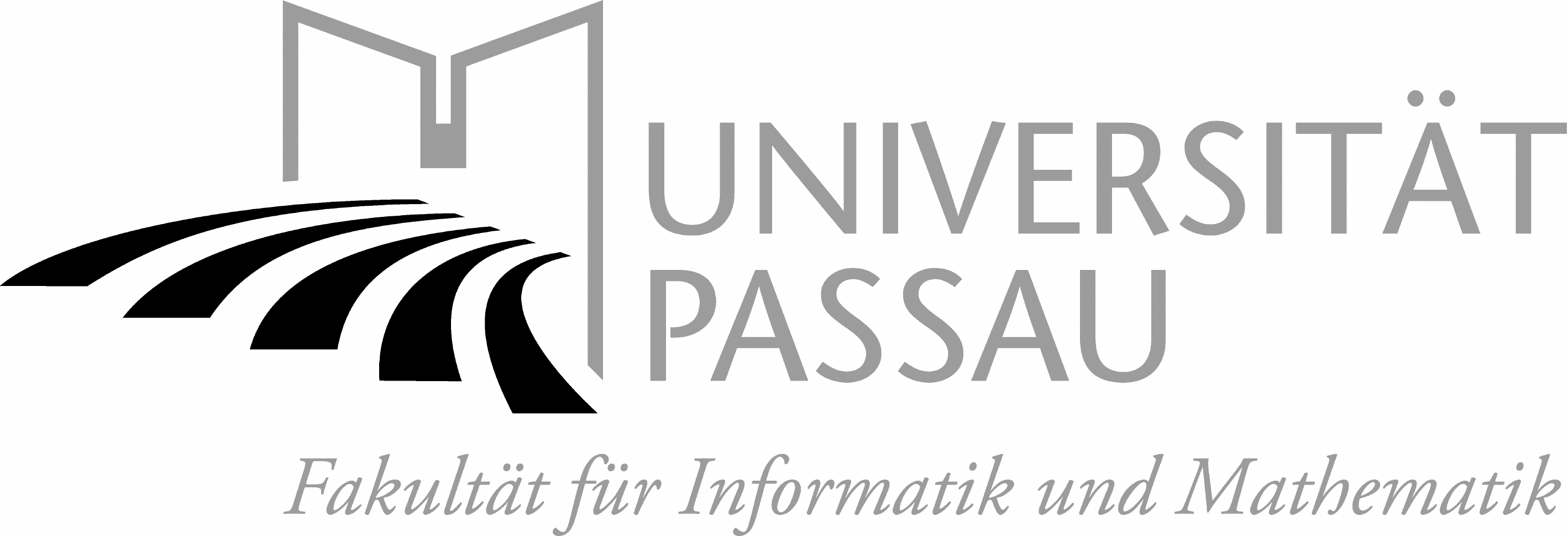} \\[1cm]
Technical Report, Number MIP-0909\\
Department of Informatics and Mathematics\\
University of Passau, Germany\\
June 2009
\end{center}

\newpage

\pagestyle{plain}
\pagenumbering{arabic}

\title{Type-Safe Feature-Oriented Product Lines}

\author{Sven Apel$^\dagger$, Christian K{\"a}stner$^\ddagger$, Armin Gr{\"o}{\ss}linger$^\dagger$, and Christian Lengauer$^\dagger$}

\institute{$^\dagger$ Department of Informatics and Mathematics, University of Passau\\\email{$\{$apel,groesslinger,lengauer$\}$@uni-passau.de}\\
$^\ddagger$ School of Computer Science, University of Magdeburg\\\email{kaestner@iti.cs.uni-magdeburg.de}
}

\maketitle

\begin{abstract}
A \emph{feature-oriented product line} is a family of programs that share a common set of features. A \emph{feature} implements a stakeholder's requirement, represents a design decision and configuration option and, when added to a program, involves the introduction of new structures, such as classes and methods, and the refinement of existing ones, such as extending methods. With feature-oriented decomposition, programs can be generated, solely on the basis of a user's selection of features, by the composition of the corresponding feature code. A key challenge of feature-oriented product line engineering is how to guarantee the correctness of an \emph{entire} feature-oriented product line, i.e., of all of the member programs generated from different combinations of features. As the number of valid feature combinations grows progressively with the number of features, it is not feasible to check all individual programs. The only feasible approach is to have a type system check the entire code base of the feature-oriented product line. We have developed such a type system on the basis of a formal model of a feature-oriented Java-like language. We demonstrate that the type system ensures that \emph{every} valid program of a feature-oriented product line is well-typed and that the type system is complete.
\end{abstract}

\section{Introduction}
\emph{Feature-oriented programming} (\emph{FOP}) aims at the modularization of programs in terms of features~\cite{Pre97,BSR04}. A \emph{feature} implements a stakeholder's requirement and is typically an increment in program functionality~\cite{Pre97,BSR04}. Contemporary feature-oriented programming languages and tools such as AHEAD~\cite{BSR04}, Xak~\cite{ADT07}, CaesarJ~\cite{MO04}, Classbox/J~\cite{BDN05}, FeatureHouse~\cite{AKL09fh}, and FeatureC++~\cite{ALR+05b} provide a variety of mechanisms that support the specification, modularization, and composition of features. A key idea is that a feature is implemented by a distinct code unit, called a \emph{feature module}. When added to a base program, it introduces new structures, such as classes and methods, and refines existing ones, such as extending methods~\cite{LBC05,ALS08tse}. A program that is decomposed into features is called henceforth a \emph{feature-oriented program}.\footnote{Typically, feature-oriented decomposition is orthogonal to class-based or functional decomposition~\cite{TOH+99,MLW+01,SB02}. A multitude of modularization and composition mechanisms~\cite{BC90,FKF98,DNS+06,MM89,MK03,RAB+92,VN96} have been developed in order to allow programmers to decompose a program along multiple dimensions~\cite{TOH+99}. Feature-oriented languages and tools provide a significant subset of these mechanisms~\cite{ALS08tse}.}

Beside the decomposition of programs into features, the concept of a feature is useful for distinguishing different, related programs thus forming a \emph{software product line}~\cite{KCH+90,CE00}. Typically, programs of a common domain share a set of features but also differ in other features. For example, suppose an email client for mobile devices that supports the protocols IMAP and POP3 and another client that supports POP3, MIME, and SSL encryption. With a decomposition of the two programs into the features \feature{IMAP}, \feature{POP3}, \feature{MIME}, and \feature{SSL}, both programs can share the code of the feature \feature{POP3}. Since mobile devices have only limited resources, unnecessary features should be removed.

With feature-oriented decomposition, programs can be generated solely on the basis of a user's selection of features by the composition of the corresponding feature modules. Of course, not all combinations of features are legal and result in correct programs~\cite{Bat05}. A \emph{feature model} describes which features can be composed in which combinations, i.e., which programs are \emph{valid}~\cite{KCH+90,CE00}. It consists of an (ordered) set of features and a set of constraints on feature combinations~\cite{CE00,Bat05}. For example, our email client may have different rendering engines for HTML text, e.g., the Mozilla engine or the Safari engine, but only one at a time. A set of feature modules along wit a feature model is called a \emph{feature-oriented product line}~\cite{Bat05}. 

An important question is how the correctness of feature-oriented programs, in particular, and product lines, in general, can be guaranteed. A first problem is that contemporary feature-oriented languages and tools usually involve a code generation step during composition in which the code is transformed into a lower-level representation. In previous work, we have addressed this problem by modeling feature-oriented mechanisms directly in the formal syntax and semantics of a core language, called \emph{Feature Featherweight Java} (\emph{FFJ}). The type system of \ffj ensures that the composition of feature modules is type-safe~\cite{AKL08ffj}.

In this paper, we address a second problem: How can the correctness of an \emph{entire} feature-oriented product line be guaranteed? A naive approach would be to type-check all valid programs of a product line using a type checker like the one of \ffj~\cite{AKL08ffj}. However, this approach does not scale; already for 34 implemented optional features, a variant can be generated for every person on the planet. Noticing this problem, Czarnecki and Pietroszek~\cite{CP06} and Thaker et al.~\cite{TBK+07} suggested the development of a type system that checks the entire code base of the feature-oriented product line, instead of all individual feature-oriented programs. In this scenario, a type checker must analyze \emph{all} feature modules of a product line on the basis of the feature model. We will show that, with this information, the type checker can ensure that \emph{every} valid program variant that can be generated is type-safe. Specifically, we make the following contributions:
\begin{itemize}
\item We provide a condensed version of \ffj, which is in many respects more elegant and concise than its predecessor~\cite{AKL08ffj}.
\item We develop a formal type system that uses information about features and constraints on feature combinations in order to type-check a product line without generating every program.
\item We prove \emph{correctness} by proving that every program generated from a well-formed product line is well-formed, as long as the feature selection satisfies the constraints of the product line. Furthermore, we prove \emph{completeness} by proving that the well-typedness of all programs of a product line guarantees that the product line is well-typed as a whole.
\item We offer an implementation of \ffj, including the proposed type system, which can be downloaded for evaluation and for experiments with further feature-oriented language and typing mechanisms.
\end{itemize}

Or work differs in many respects from previous and related work (see Section~\ref{sec:related_work} for a comprehensive discussion). Most notably, Thaker et al.\ have implemented a type system for feature-oriented product lines and conducted several case studies~\cite{TBK+07}. We take their work further with a formalization and a correctness and completeness proof. 

Furthermore, our work differs in many respects from previous work on modeling and type-checking feature-oriented and related programming mechanisms. Most notably, we model the feature-related mechanisms directly in \ffj's syntax and semantics, without any transformation to a lower-level representation, and we stay very close to the syntax of contemporary feature-oriented languages and tools (see Section~\ref{sec:related_work}). We begin with a brief introduction to \ffj.

\section{Feature-Oriented Programs in \ffj}
\label{sec:ffj}
In this section, we introduce the language \ffj. Originally, \ffj was designed for feature-oriented programs~\cite{AKL08ffj,AKL08ffjtr}. We extend \ffj in Section~\ref{sec:ffjpl} to support feature-oriented product lines, i.e., to support the representation of multiple alternative program variants at a time.

\subsection{An Overview of \ffj}
\ffj is a lightweight feature-oriented language that has been inspired by \emph{Featherweight Java} (\emph{FJ})~\cite{IPW01}. As with \fj, we have aimed at minimality in the design of \ffj. \ffj provides basic constructs like classes, fields, methods, and inheritance and only a few new constructs capturing the core mechanisms of feature-oriented programming. But, so far, \ffj's type system has not supported the development of feature-oriented product lines. That is, the feature modules written in \ffj are interpreted as a single program. We will change this in Section~\ref{sec:ffjpl}.

An \ffj program consists of a set of classes and refinements. A \emph{refinement} extends a class that has been introduced previously. Each class and refinement is associated with a feature. We say that a feature \emph{introduces} a class or \emph{applies} a refinement to a class. Technically, the mapping between classes/refinements and the features they belong to can be established in different ways, e.g., by extending the language with modules representing features~\cite{MO04,BDN05,DCB09} or by grouping classes and refinements that belong to a feature in packages or directories~\cite{BSR04,ALR+05b}.

Like in \fj, each class declares a superclass, which may be the class \code{Object}. Refinements are defined using the keyword \code{refines}. The semantics of a refinement applied to a class is that the refinement's members are added to and merged with the member of the refined class. This way, a refinement can \emph{add} new fields and methods to the class and \emph{override} existing methods (declared by \code{overrides}). 

\begin{figure}[tbh]
\centering
\begin{minipage}{4.2cm}
\fheader{EmailClient}
\begin{lstlisting}[firstnumber=1,frame=top|bottom,basicstyle=\sf\scriptsize,
keywordstyle=\sf\scriptsize\bfseries,commentstyle=\sf\scriptsize\it]{ar}
class Msg extends Object { 
  String serialize() { ... }
}
class Trans extends Object {
  $\mbox{\sf Bool}$ send(Msg m) { ... }
}
\end{lstlisting}
\fheader{SSL}
\begin{lstlisting}[frame=top|bottom,basicstyle=\sf\scriptsize,
keywordstyle=\sf\scriptsize\bfseries,commentstyle=\sf\scriptsize\it]{ar}
class SSL extends Object {
  Trans trans;
  $\mbox{\sf Bool}$ send(Msg m) { ... }
}
refines class Trans {
  Key key;
  overrides Bool send(Msg m) { 
    return new SSL(this).send(m);
  }
}
\end{lstlisting}
\vspace{-2ex}
\end{minipage}
\hspace{0.5cm}
\begin{minipage}{6.5cm}
\centering
\includegraphics[scale=0.6,angle=0]{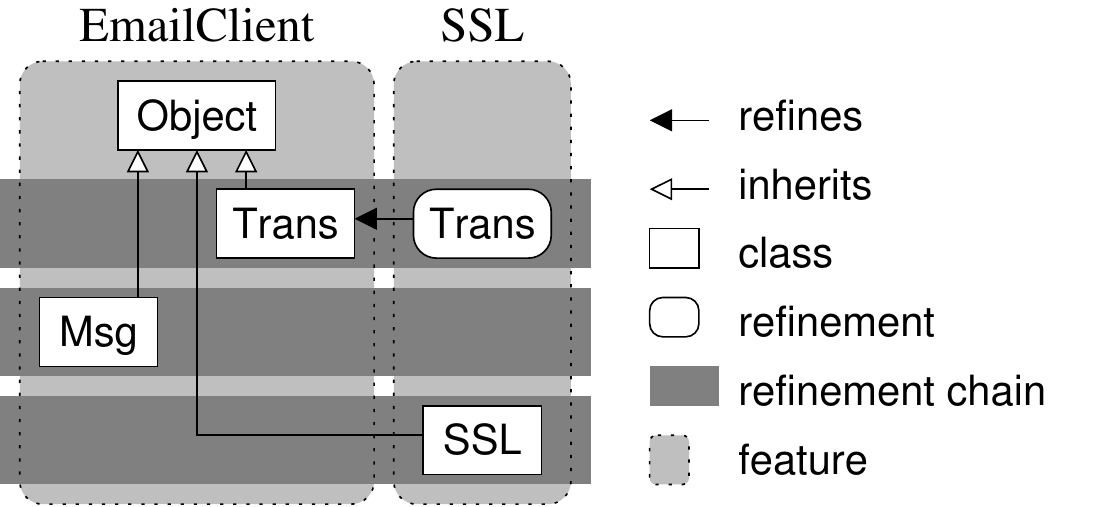}
\end{minipage}
\caption{A feature-oriented email client supporting SSL encryption.}
\label{fig:mt}
\end{figure}

On the left side in Figure~\ref{fig:mt}, we show an excerpt of the code of a basic email client, called \feature{EmailClient}, (top) and a feature, called \feature{SSL}, (bottom) in \ffj. The feature \feature{SSL} adds the class \code{SSL} (Lines~7--10) to the email client's code base and refines the class \code{Trans} in order to encrypt outgoing messages (Lines~11--15). To this effect, the refinement of \code{Trans} adds a new field \code{key} (Line~12) and overrides the method \code{send} of class \code{Trans} (Lines~13-15).

Typically, a programmer applies multiple refinements to a class by composing a sequence of features. This is called a \emph{refinement chain}.  A refinement that is applied immediately before another refinement in the chain is called its \emph{predecessor}. The order of the refinements in a refinement chain is determined by their composition order. On the right side in Figure~\ref{fig:mt}, we depict the refinement and inheritance relationships of our email example. 

Fields are unique within the scope of a class and its inheritance hierarchy and refinement chain. That is, a refinement or subclass is not allowed to add a field that has already been defined in a predecessor in the refinement chain or in a superclass. For example, a further refinement of \code{Trans} would not be allowed to add a field \code{key}, since \code{key} has been introduced by a refinement of feature \feature{SSL} already. With methods, this is different. A refinement or subclass may add new methods (overloading is prohibited) \emph{and} override existing methods. In order to distinguish the two cases, \ffj expects the programmer to declare whether a method overrides an existing method (using the modifier \code{overrides}). For example, the refinement of \code{Trans} in feature \feature{SSL} overrides the method \code{send} introduced by feature \feature{Mail}; for subclasses, this is similar.

The distinction between method introduction and overriding allows the type system to check (1) whether an introduced method inadvertently replaces or occludes an existing method with the same name and (2) whether, for every overriding method, there is a proper method to be overridden. Apart from the modifier \code{overrides}, a method in \ffj is similar to a method in \fj. That is, a method body is an expression (prefixed with \textsf{return}) and not a sequence of statements. This is due to the functional nature of \ffj and \fj. Furthermore, overloading of methods (introducing methods with equal names and different argument types) is not allowed in \ffj (and \fj).

\begin{figure}[tbh]
\centering
\includegraphics[scale=0.6,angle=0]{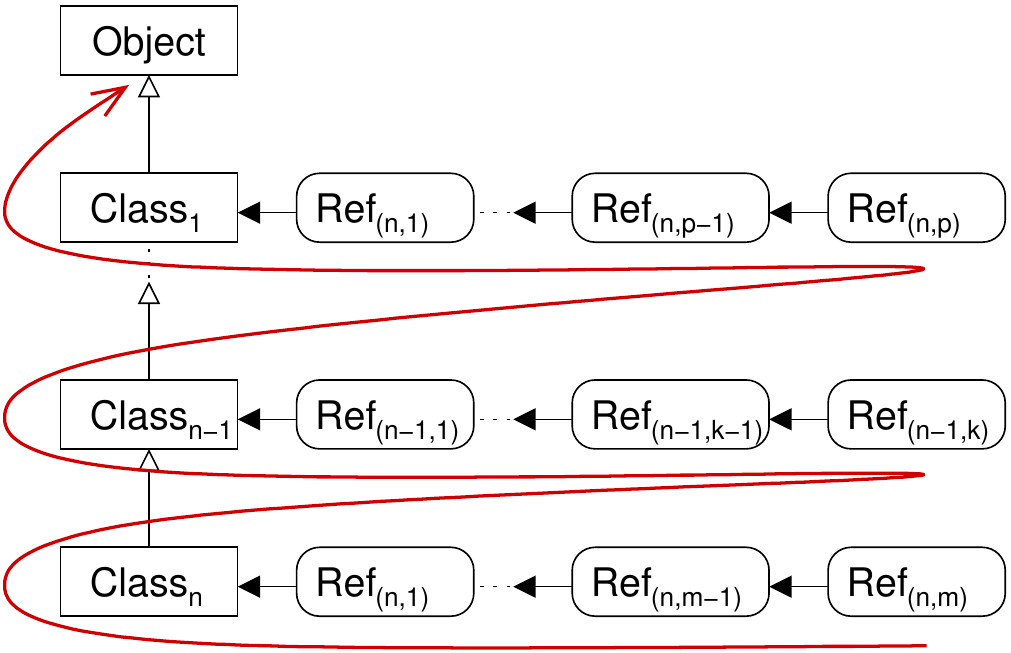}
\caption{Order of method body and field lookup in \ffj.}
\label{fig:mfl}
\end{figure}

As shown in Figure~\ref{fig:mt}, refinement chains grow from left to right and inheritance hierarchies from top to bottom. When looking up a method body, \ffj traverses the combined inheritance and refinement hierarchy of an object and selects the right-most and bottom-most body of a method declaration or method refinement that is compatible. This kind of lookup is necessary since we model features \emph{directly} in \ffj, instead of generating and evaluating \fj code~\cite{LSZ09}. First, the \ffj calculus looks for a method declaration in the refinement chain of the object's class, starting with the last refinement back to the class declaration itself. The first body of a matching method declaration is returned. If the method is not found in the class' refinement chain or in its own declaration, the methods in the superclass (and then the superclass' superclass, etc.) are searched, each again from the most specific refinement of the class declaration itself. The field lookup works similarly, except that the entire inheritance and refinement hierarchy is searched and the fields are accumulated in a list. In Figure~\ref{fig:mfl}, we illustrate the processes of method body and field lookup schematically.

\subsection{Syntax of \ffj}
\label{sec:syntax}
Before we go into detail, let us explain some notational conventions. We abbreviate lists in the obvious ways: 
\begin{itemize}
\item $\mol{C}$ is shorthand for \m{C$_1$, $\ldots$\,, C$_n$} 
\item \m{\ol{C~f}} is shorthand for \m{C$_1$ f$_1$, $\ldots$ , C$_n$ f$_n$} 
\item \m{\ol{C~f};} is shorthand for \m{C$_1$ f$_1$; $\ldots$ ; C$_n$ f$_n$;} 
\item \m{\ol{t : C}} is shorthand for \m{t$_1$ : C$_1$, $\ldots$ , t$_n$ : C$_n$}
\item \m{\ol{C <: D}} is shorthand for \m{C$_1$ <: D$_1$\ $\ldots$\ C$_n$ <: D$_n$}
\item $\ldots$
\end{itemize}
Note that, depending on the context, blanks, commas, or semicolons separate the elements of a list. The context will make clear which separator is meant. The symbol $\bullet$ denotes the empty list and lists of field declarations, method declarations, and parameter names must not contain duplicates. We use the metavariables \m{A}--\m{E} for class names, \m{f}--\m{h} for field names, and \m{m} for method names. Feature names are denoted by Greek letters.

In Figure~\ref{fig:syntax}, we depict the syntax of \ffj in extended Backus-Naur-Form. An \ffj program consists of a set of class and refinement declarations. A class declaration \m{L} declares a class with the name \m{C} that inherits from a superclass \m{D} and consists of a list $\mol{C~f};$ of fields and a list $\mol{M}$ of method declarations.\footnote{The concept of a class constructor is unnecessary in \ffj and \fj~\cite{Pie02}. Its omittance simplifies the syntax, semantics, and type rules significantly without loss of generality.} A refinement declaration \m{R} consists of a list $\mol{C~f};$ of fields and a list $\mol{M}$ of method declarations. 
\begin{figure*}
\begin{center}
\begin{tabular}{c@{\qquad}|@{\qquad}c}
\begin{tabular}{l l}
\textsf{L} ::= & \hfill \textit{class declarations:} \\
\multicolumn{2}{l}{
\quad \textsf{class C extends D \{ \ol{C f}; \ol{M} \}}
} \\[2ex]
\textsf{R} ::= & \hfill \textit{refinement declarations:} \\
\multicolumn{2}{l}{
\quad \textsf{refines class C \{ \ol{C f}; \ol{M} \}}
} \\[2ex]
\textsf{M} ::= & \hfill \textit{method declarations:} \\
\multicolumn{2}{l}{
\quad \textsf{[overrides] C m(\ol{C x}) \{ return t; \}}
}
\end{tabular}
&
\begin{tabular}{l l }
\textsf{t} ::= & \hfill \textit{terms:} \\
\multicolumn{2}{l}{
\quad \textsf{x}  \hfill \textit{variable}
}\\
\multicolumn{2}{l}{
\quad \textsf{t.f}  \hfill \textit{field access}
} \\
\multicolumn{2}{l}{
\quad \textsf{t.m(\ol{t})}  \hfill \quad \textit{method invocation}
} \\
\multicolumn{2}{l}{
\quad \textsf{new C(\ol{t})}  \hfill \textit{object creation}
} \\
\multicolumn{2}{l}{
\quad \textsf{(C)\,t}  \hfill \textit{cast}
} \\[2ex]
\textsf{v} ::= & \hfill \textit{values:} \\
\multicolumn{2}{l}{
\quad \textsf{new C(\ol{v})} \hfill \textit{object creation}
}
\end{tabular}
\end{tabular}
\end{center}
\caption{Syntax of \ffj in extended BNF.}
\label{fig:syntax}
\end{figure*}

A method \m{m} expects a list \m{\ol{C~x}} of arguments and declares a body that returns only a single expression \m{t} of type \m{C}. Using the modifier \code{overrides}, a method declares that it intends to override another method with the same name and signature. 
Where we want to distinguish methods that override others and methods that do not override others, we call the former \emph{method introductions} and the latter \emph{method refinements}

Finally, there are five forms of terms: the variable, field access, method invocation, object creation, and type cast, which are taken from \fj without change. The only values are object creations whose arguments are values as well.

\subsection{\ffj's Class Table}

Declarations of classes and refinements can be looked up via a class table $\mait{CT}$. The compiler fills the class table during the parser pass. In contrast to \fj, class and refinement declarations are identified not only by their names but, additionally, by the names of the enclosing features. For example, in order to retrieve the declaration of class \code{Trans}, introduced by feature \feature{Mail}, in our example of Figure~\ref{fig:mt}, we write $\mait{CT}($\textsc{Mail}$\masf{.Trans})$; in order to retrieve the refinement of class \code{Trans} applied by feature \feature{SSL}, we write $\mait{CT}($\textsc{SSL}$\masf{.Trans})$. We call \qt{\Phi}{C} the \emph{qualified type} of class $\masf{C}$ in feature $\man{\Phi}$. In \ffj, class and refinement declarations are unique with respect to their qualified types. This property is ensured because of the following sanity conditions: a feature is not allowed 
\begin{itemize}
\item to introduce a class or refinement twice inside a single feature module and
\item to refine a class that the feature has just introduced.
\end{itemize}
These are common sanity conditions in feature-oriented languages and tools~\cite{BSR04,ALR+05b,AKL09fh}.

As for \fj, we impose further sanity conditions on the class table and the inheritance relation:
\begin{itemize}
\item $\mathit{CT}(\qtm{\Phi}{C}) = \masf{class C\ldots}$ or $\masf{refines class C\ldots}$ for every qualified type $\qtm{\Phi}{C} \in \mathit{dom}(\mathit{CT})$; Feature $\man{Base}$ plays the same role for features as $\masf{Object}$ plays for classes; it is a symbol denoting the empty feature at which lookups terminate.
\item $\qtm{Base}{Object} \notin \mathit{dom}(\mathit{CT})$; 
\item for every class name \m{C} appearing anywhere in $\mathit{CT}$, we have $\qtm{\Phi}{C} \in \mathit{dom}(\mathit{CT})$ for at least one feature $\man{\Phi}$; and 
\item the inheritance relation contains no cycles (incl.\ self-cycles).
\end{itemize}

\subsection{Refinement in \ffj}
\label{sec:ref_ffj}
Information about the refinement chain of a class can be retrieved using the refinement table $\mathit{RT}$. The compiler fills the refinement table during the parser pass. $\mathit{RT}(\masf{C})$ yields a list of all features that either introduce or refine class $\masf{C}$. The leftmost element of the result list is the feature that introduces the class $\masf{C}$ and, then, from left to right, the features are listed that refine class $\masf{C}$ in the order of their composition. In our example of Figure~\ref{fig:mt}, $\mathit{RT}(\masf{Trans})$ yields the list $\mbox{\textsc{EmailClient}},\mbox{\textsc{SSL}}$. There is only a single sanity condition for the refinement table:
\begin{itemize}
\item $\mathit{RT}(\masf{C}) = \overline{\man{\Phi}}$ for every type $\masf{C} \in \mathit{dom}(\mathit{CT})$, with $\overline{\man{\Phi}}$ being the features that introduce and refine class $\masf{C}$.
\end{itemize}

In Figure~\ref{fig:ref}, we show two functions for the navigation of the refinement chain that rely on $\mathit{RT}$. Function $\mathit{last}$ returns, for a class name \m{C}, a qualified type $\man{\Psi}_n.\masf{C}$, in which $\man{\Psi}_n$ refers to the feature that applies the final refinement to class \m{C}; if a class is not refined at all, $\man{\Psi}_n$ refers to the feature that introduces class \m{C}. Function $\mathit{pred}$ returns, for a qualified type \qt{\Phi}{C}, another qualified type $\man{\Psi}_n.\masf{C}$, in which $\man{\Psi}_n$ refers to the feature that introduces or refines class $\masf{C}$ and that is the immediate predecessor of $\man{\Phi}$ in the refinement chain; if there is no predecessor, \qt{Base}{Object} is returned.
\begin{figure*}
\begin{center}
\begin{tabular}{l l}
\textit{Navigating along the refinement chain} &  \\[2ex]
\multicolumn{2}{c}{
$\dfrac{
\mathit{RT}(\masf{C}) = \overline{\man{\Psi}}
}
{
\last{C} = \man{\Psi}_n.\masf{C}
}$
\qquad
$\dfrac{
\mathit{RT}(\masf{C}) = \overline{\man{\Psi}}, \man{\Phi}, \overline{\man{\Omega}} \qquad \overline{\man{\Psi}} \neq \bullet
}
{
\pred{C}{\Phi} = \man{\Psi}_n.\masf{C} 
}$ 
\qquad
$\dfrac{
\mathit{RT}(\masf{C}) = \man{\Phi}, \overline{\man{\Omega}}
}
{
\pred{C}{\Phi} = \man{Base}.\masf{Object}
}$ 
}
\end{tabular}
\end{center}
\caption{Refinement in \ffj.}
\label{fig:ref}
\end{figure*}

\subsection{Subtyping in \ffj}
In Figure~\ref{fig:sub}, we show the subtype relation of \ffj. The subtype relation $\masf{<:}$ is defined by one rule each for reflexivity and transitivity and one rule for relating the type of a class to the type of its immediate superclass. It is not necessary to define subtyping over qualified types because only classes (not refinements) declare superclasses and there is only a single declaration per class.
\begin{figure*}
\begin{center}
\begin{tabular}{l l}
\textit{Subtyping} & \hfill \fbox{$\masf{C}\, <:\, \masf{D}$} \\[2ex]
\multicolumn{2}{c}{
$\masf{C}\, <:\, \masf{C}$ \qquad
$\dfrac{\masf{C}\, <:\, \masf{D} \qquad \masf{D}\, <:\, \masf{E}}{\masf{C}\, <:\, \masf{E}}$
\qquad
$\dfrac{
\mait{CT}(\qtm{\Phi}{C}) = \masf{class C extends D \{ \ldots\}} 
}
{\masf{C}\, <:\, \masf{D}
}$
}
\end{tabular}
\end{center}
\caption{Subtyping in \ffj.}
\label{fig:sub}
\end{figure*}

\subsection{Auxiliary Definitions of \ffj}
In Figure~\ref{fig:aux}, we show the auxiliary definitions of \ffj. Function $\mait{fields}$ searches the refinement chain from right to left and accumulates the fields into a list (using the comma as concatenation operator). If there is no further predecessor in the refinement chain, i.e., we have reached a class declaration, then the refinement chain of the superclass is searched (see~Figure~\ref{fig:mfl}). If \qt{Base}{Object} is reached, the empty list is returned (denoted by $\bullet$).
\begin{figure*}
\begin{center}
\begin{tabular}{c c}
\multicolumn{1}{l}{\textit{Field lookup}} \hfill & \hfill \fbox{$\fieldsX{C}{\Phi} = \mol{C~f}$} \\[2ex]
\multicolumn{2}{c}{
$\fieldsX{Object}{Base} = \bullet$
}\\[2ex]
\multicolumn{2}{c}{
$\dfrac{
\mathit{CT}(\qtm{\Phi}{C}) = \masf{class C extends D \{\ } \mol{C~f};\ \mol{M}\ \masf{\}}
}
{\fieldsX{C}{\Phi} = \fieldsY{\last{D}},\, \mol{C~f}}$
\quad
$\dfrac{
\mathit{CT}(\qtm{\Phi}{C}) = \masf{refines class C \{\ } \mol{C~f};\ \mol{M}\ \masf{\}}
}
{\fieldsX{C}{\Phi} = \fieldsY{\pred{C}{\Phi}},\, \mol{C~f}}$
}\\[4ex]
\multicolumn{1}{l}{\textit{Method body lookup}} \hfill & \hfill \fbox{$\mbodyA{m}{C}{\Phi} = (\mol{x}, \masf{t})$} \\[2ex]
$\dfrac{
\begin{array}{c}
\masf{[overrides] B m(\ol{B~x}) \{ return t; \}} \in \mol{M} \\
\mathit{CT}(\qtm{\Phi}{C}) = \masf{class C extends D \{\ } \mol{C~f};\ \mol{M}\ \masf{\}}
\end{array}
}
{\mbodyA{m}{C}{\Phi} = (\mol{x}, \masf{t})}$
&
$\dfrac{
\begin{array}{c}
\masf{m} \mbox{ is not defined in } \mol{M}\\
\mathit{CT}(\qtm{\Phi}{C}) = \masf{class C extends D \{\ } \mol{C~f};\ \mol{M}\ \masf{\}}
\end{array}
}
{\mbodyA{m}{C}{\Phi} = \mbodyB{m}{\last{D}}}$
\\[4ex]
$\dfrac{
\begin{array}{c}
\masf{[overrides] B m(\ol{B~x}) \{ return t; \}} \in \mol{M} \\
\mathit{CT}(\qtm{\Phi}{C}) = \masf{refines class C \{\ } \mol{C~f};\ \mol{M}\ \masf{\}}
\end{array}
}
{\mbodyA{m}{C}{\Phi} = (\mol{x}, \masf{t})}$
&
$\dfrac{
\begin{array}{c}
\masf{m} \mbox{ is not defined in } \mol{M}\\
\mathit{CT}(\qtm{\Phi}{C}) = \masf{refines class C \{\ } \mol{C~f};\ \mol{M}\ \masf{\}}
\end{array}
}
{\mbodyA{m}{C}{\Phi} = \mbodyB{m}{\pred{C}{\Phi}}}$
\\[4ex]
\multicolumn{1}{l}{\textit{Method type lookup}} \hfill & \hfill \fbox{$\mtypeX{m}{C}{\Phi} = \mol{C} \fct \masf{C}$} \\[2ex]
\multicolumn{2}{c}{
$\dfrac{
\begin{array}{c}
\masf{B$_0$ m(\ol{B~x}) \{ return t; \}} \in \mol{M} \\
\mathit{CT}(\qtm{\Phi}{C}) = \masf{class C extends D \{\ } \mol{C~f};\ \mol{M}\ \masf{\}}
\end{array}
}
{\mtypeX{m}{C}{\Phi} = \mol{B} \fct \masf{B$_0$}}$
\quad
$\dfrac{
\begin{array}{c}
\masf{m} \mbox{ is not defined in } \mol{M}\\
\mathit{CT}(\qtm{\Phi}{C}) = \masf{class C extends D \{\ } \mol{C~f};\ \mol{M}\ \masf{\}}
\end{array}
}
{\mtypeX{m}{C}{\Phi} = \mtypeY{m}{\last{D}}}$
}\\[4ex]
\multicolumn{2}{c}{
$\dfrac{
\begin{array}{c}
\masf{B$_0$ m(\ol{B~x}) \{ return t; \}} \in \mol{M} \\
\mathit{CT}(\qtm{\Phi}{C}) = \masf{refines class C \{\ } \mol{C~f};\ \mol{M}\ \masf{\}}
\end{array}
}
{\mtypeX{m}{C}{\Phi} = \mol{B} \fct \masf{B$_0$}}$
\quad
$\dfrac{
\begin{array}{c}
\masf{m} \mbox{ is not defined in } \mol{M} \\
\mathit{CT}(\qtm{\Phi}{C}) = \masf{refines class C \{\ } \mol{C~f};\ \mol{M}\ \masf{\}}
\end{array}
}
{\mtypeX{m}{C}{\Phi} = \mtypeY{m}{\pred{C}{\Phi}}}$
}\\[4ex]
\multicolumn{1}{l}{\textit{Valid class introduction}} \hfill & \hfill \fbox{$\mathit{introduce}(\qtm{\Phi}{C})$} \\[2ex]
\multicolumn{2}{c}{
$\dfrac{
\nexists\, \man{\Psi}\, :\, (\mathit{CT}(\qtm{\Psi}{C}) = \masf{class C}\ldots\ \wedge\ \man{\Phi} \ne \man{\Psi})
}
{\mathit{introduce}(\qtm{\Phi}{C})}$
}\\[4ex]
\multicolumn{1}{l}{\textit{Valid field introduction}} \hfill & \hfill \fbox{$\introduceX{f}{C}{\Phi}$} \\[2ex]
\multicolumn{2}{c}{
$\dfrac{
\mathit{fields}(\qtm{\Phi}{C}) = \mol{E~h} \qquad \masf{f} \notin \mol{h}
}
{\introduceX{f}{C}{\Phi}}$
}\\[4ex]
\multicolumn{1}{l}{\textit{Valid method introduction}} \hfill & \hfill \fbox{$\introduceX{m}{C}{\Phi}$} \\[2ex]
\multicolumn{2}{c}{
$\dfrac{
(\masf{m},\qtm{\Phi}{C}) \notin \mathit{dom}(\mathit{mtype})
}
{\introduceX{m}{C}{\Phi}}$
}\\[4ex]
\multicolumn{1}{l}{\textit{Valid class refinement}} \hfill & \hfill \fbox{$\mathit{refine}(\qtm{\Phi}{C})$} \\[2ex]
\multicolumn{2}{c}{
$\dfrac{
\mathit{RT}(\masf{C}) = \overline{\man{\Psi}}, \man{\Phi}, \overline{\man{\Omega}} \qquad \mathit{CT}(\man{\Psi}_1.\masf{C}) = \masf{class C}\ldots
}
{\mathit{refine}(\qtm{\Phi}{C})}$
}\\[4ex]
\multicolumn{1}{l}{\textit{Valid method overriding}} \hfill & \hfill \fbox{$\overrideX{m}{C}{\Phi}{\mol{C} \fct \masf{C$_0$}}$} \\[2ex]
\multicolumn{2}{c}{
$\dfrac{
\mtypeX{m}{C}{\Phi} = \mol{B} \fct \masf{B$_0$} \qquad  \mol{C} = \mol{B} \qquad \quad \masf{C$_0$} = \masf{B$_0$}
}
{\overrideX{m}{C}{\Phi}{\mol{C} \fct \masf{C$_0$}}}$
}
\end{tabular}
\end{center}
\caption{Auxiliary definitions of \ffj.}
\label{fig:aux}
\end{figure*}

Function $\mait{mbody}$ looks up the most specific and most refined body of a method $\masf{m}$. A body consists of the formal parameters $\mol{x}$ of a method and the actual term \m{t} representing the content. The search is like in $\mait{fields}$. First, the refinement chain is searched from right to left and, then, the superclasses' refinement chains are searched, as illustrated in Figure~\ref{fig:mfl}. Note that $\masf{[overrides]}$ means that a given method declaration may (or may not) have the modifier. This way, we are able to define uniform rules for method introduction and method refinement. Function $\mathit{mtype}$ yields the signature $\mol{B} \fct \masf{B$_0$}$ of a declaration of method $\masf{m}$. The lookup is like in $\mathit{mbody}$.

Predicate $\mathit{introduce}$ is used to check whether a class has been introduced by multiple features and whether a field or method has been introduced multiple times in a class. Precisely, it states, in the case of classes, whether $\masf{C}$ has not been introduced by any feature other than $\man{\Phi}$ and whether a method $\masf{m}$ or a field \m{f} has not been introduced by \qt{\Phi}{C} or in any of its predecessors or superclasses. To evaluate it, we check, in the case of classes, whether $\mathit{CT}(\qtm{\Psi}{C})$ yields a class declaration or not, for any feature $\man{\Psi}$ different from $\man{\Phi}$, in the case of methods, whether $\mathit{mtype}$ yields a signature or not and, in the case of fields, whether \m{f} is defined in the list of fields returned by $\mathit{fields}$. 

Predicate $\mathit{refine}$ states whether, for a given refinement, a proper class has been declared previously in the refinement chain. The predicate $\mathit{override}$ states whether a method $\masf{m}$ has been introduced before in some predecessor of \qt{\Phi}{C} and whether the previous declaration of $\masf{m}$ has the given signature.

\subsection{Evaluation of \ffj Programs}
\label{sec:eval}
Each \ffj program consists of a class table and a term.\footnote{The refinement table is not relevant for evaluation.} The term is evaluated using the evaluation rules shown in Figure~\ref{fig:eval}. The evaluation terminates when a value, i.e., a term of the form $\masf{new C(\ol{v})}$, is reached. Note that we use a \emph{direct semantics} of class refinement~\cite{LSZ09}. That is, the field and method lookup mechanisms incorporate all refinements when a class is searched for fields and methods. An alternative, which is discussed in Section~\ref{sec:related_work}, would be a \emph{flattening semantics}, i.e., to merge a class in a preprocessing step with all of its refinements into a single declaration.
\begin{figure*}
\begin{center}
\begin{tabular}{c r}
$\dfrac{
\fieldsY{\last{C}} = \mol{C~f}
}{
\masf{(new C(\ol{v})).}\masf{f$_i$}\ \longrightarrow\ \masf{v$_i$}
}$ 
& 
$
(\mbox{\textsc{E-ProjNew}})
$
\\[4ex]
$\dfrac{
\mbodyB{m}{\last{C}} = (\mol{x}, \masf{t$_0$})
}{
\masf{(new C(\ol{v})).}\masf{m(\ol{u})}\ \longrightarrow\ \left[ \overline{\masf{x} \mapsto \masf{u}}, \masf{this} \mapsto \masf{new C(\ol{v})} \right] \masf{t$_0$}
}$
& 
$
(\mbox{\textsc{E-InvkNew}})
$
\\[4ex]
$\dfrac{
\masf{C}\, <:\, \masf{D}
}{
\masf{(D)(new C(\ol{v}))}\ \longrightarrow\ \masf{new C(\ol{v})}
}$
& 
$
(\mbox{\textsc{E-CastNew}})
$
\\[4ex]
$\dfrac{
\masf{t$_0$}\ \longrightarrow\ \masf{t$'_0$}
}{
\masf{t$_0$}\masf{.f}\ \longrightarrow\ \masf{t$'_0$}\masf{.f}
}$
& 
$
(\mbox{\textsc{E-Field}})
$
\\[4ex]
$\dfrac{
\masf{t$_0$}\ \longrightarrow\ \masf{t$'_0$}
}{
\masf{t$_0$}\masf{.m(\ol{t})}\ \longrightarrow\ \masf{t$'_0$}\masf{.m(\ol{t})}
}$ 
& 
$
(\mbox{\textsc{E-InvkRecv}})
$
\\[4ex]
$\dfrac{
\masf{t$_i$}\ \longrightarrow\ \masf{t$'_i$}
}{
\masf{v$_0$}\masf{.m(\ol{v}, t$_i$, \ol{t})}\ \longrightarrow\ \masf{v$_0$}\masf{.m(\ol{v}, t$'_i$, \ol{t})}
}$
& 
$
(\mbox{\textsc{E-InvkArg}})
$
\\[4ex]
$\dfrac{
\masf{t$_i$}\ \longrightarrow\ \masf{t$'_i$}
}{
\masf{new C(\ol{v}, t$_i$, \ol{t})}\ \longrightarrow\ \masf{new C(\ol{v}, t$'_i$, \ol{t})}
}$ 
& 
$
(\mbox{\textsc{E-NewArg}})
$
\\[4ex]
$\dfrac{
\masf{t$_0$}\ \longrightarrow\ \masf{t$'_0$}
}{
\masf{(C)t$_0$.f}\ \longrightarrow\ \masf{(C)t$'_0$.f}
}$
& 
$
(\mbox{\textsc{E-Cast}})
$
\end{tabular}
\end{center}
\caption{Evaluation of \ffj programs.}
\label{fig:eval}
\end{figure*}
 
Using the subtype relation $<:$ and the auxiliary functions $\mait{fields}$ and $\mait{mbody}$, the evaluation of \ffj is fairly simple. The first three rules are most interesting (the remaining rules are just congruence rules). Rule \textsc{E-ProjNew} describes the projection of a field from an instantiated class. A projected field $\masf{f$_i$}$ evaluates to a value $\masf{v$_i$}$ that has been passed as argument to the instantiation. Function $\mait{fields}$ is used to look up the fields of the given class. It receives $\last{C}$ as argument since we want to search the entire refinement chain of class $\masf{C}$ from right to left (cf.~Figure~\ref{fig:mfl}).
 
Rule \textsc{E-ProjInvk} evaluates a method invocation by replacing the invocation with the method's body. The formal parameters of the method are substituted in the body for the arguments of the invocation; the value on which the method is invoked is substituted for \m{this}. The function $\mait{mbody}$ is called with the last refinement of the class \m{C} in order to search the refinement chain from right to left and return the most specific method body (cf.~Figure~\ref{fig:mfl}).

Rule \textsc{E-CastNew} evaluates an upcast by simply removing the cast. Of course, the premise must be that the cast is really an upcast and not a downcast or an incorrect cast.

\subsection{Type Checking \ffj Programs}
The type relation of \ffj consists of the type rules for terms and the well-formedness rules for classes, refinements, and methods, shown in Figures~\ref{fig:type1} and~\ref{fig:type2}.
\begin{figure*}
\begin{center}
\begin{tabular}{l l}
\textit{Term typing} & \hfill \fbox{$\man{\Gamma}\ \vdash\ \masf{t : C}$} \\[2ex]
\multicolumn{2}{c}{
\hfill 
$\dfrac{\masf{x : C}\ \in\ \man{\Gamma}}{\man{\Gamma}\ \vdash\ \masf{x : C}}$ 
\hfill
$
(\mbox{\sc T-Var})
$
}\\[4ex]
\multicolumn{2}{c}{
\hfill 
$\dfrac{
\man{\Gamma}\ \vdash\ \masf{t$_0$}\ \masf{: C$_0$} \qquad \fieldsY{\last{C$_0$}} = \mol{C~f} 
}
{
\man{\Gamma}\ \vdash\ \masf{t$_0$} \masf{.f$_i$}\ \masf{: C$_i$}
}$
\hfill
$
(\mbox{\sc T-Field})
$
}\\[4ex]
\multicolumn{2}{c}{
\hfill 
$\dfrac{
\man{\Gamma}\ \vdash\ \masf{t$_0$}\ \masf{: C$_0$} \qquad \man{\Gamma}\ \vdash\ \mol{t : C} \qquad
\mtypeY{m}{\last{C$_0$}} = \mol{D} \fct \masf{C} \qquad \mol{C\, <:\, D}
}
{
\man{\Gamma}\ \vdash\ \masf{t$_0$}\masf{.m(\ol{t}) : C}
}$
\hfill 
$
(\mbox{\sc T-Invk})
$
}\\[4ex]
\multicolumn{2}{c}{
\hfill 
$\dfrac{
\man{\Gamma}\ \vdash\ \mol{t : C} \qquad \fieldsY{\last{C}} = \mol{D~f} \qquad \mol{C\, <:\, D}
}
{
\man{\Gamma}\ \vdash\ \masf{new C(\ol{t}) : C}
}$ 
\hfill 
$
(\mbox{\sc T-New})
$
}\\[4ex]
\multicolumn{2}{c}{
\hfill
$\dfrac{
\man{\Gamma}\ \vdash\ \masf{t}_0\ \masf{: D} \qquad \masf{D}\, <:\, \masf{C}
}
{
\man{\Gamma}\ \vdash\ \masf{(C)t}_0\ \masf{: C}
}$ 
\hfill 
$
(\mbox{\sc T-UCast})
$
}\\[4ex]
\multicolumn{2}{c}{
\hfill
$\dfrac{
\man{\Gamma}\ \vdash\ \masf{t}_0\ \masf{: D} \qquad \masf{C}\, <:\, \masf{D} \qquad \masf{C}\, \neq\, \masf{D}
}
{
\man{\Gamma}\ \vdash\ \masf{(C)t}_0\ \masf{: C}
}$
\hfill $
(\mbox{\sc T-DCast})
$
}\\[4ex]
\multicolumn{2}{c}{
\hfill $\dfrac{
\man{\Gamma}\ \vdash\ \masf{t}_0\ \masf{: D} \qquad \masf{C}\, \not<:\, \masf{D} \qquad \masf{D}\, \not<:\, \masf{C} \qquad
\mathit{stupid\ warning}
}
{
\man{\Gamma}\ \vdash\ \masf{(C)t}_0\ \masf{: C}
}$ \hfill ({\sc T-SCast})
}
\end{tabular}
\end{center}
\caption{Term typing in \ffj.}
\label{fig:type1}
\end{figure*}

\begin{figure*}
\begin{center}
\begin{tabular}{l l}
\textit{Method typing} & \hfill \fbox{$\masf{M} \mbox{ OK} \dashv \qtm{\Phi}{C}$} \\[1ex]
\multicolumn{2}{c}{
$\dfrac{
\begin{array}{c}
\masf{\ol{x : B},\,this : C}\ \vdash\ \masf{t$_0$}\ \masf{: E$_0$} \qquad \masf{E$_0$}\, <:\, \masf{B$_0$}\\
\mathit{CT}(\qtm{\Phi}{C}) = \masf{class C extends D \{ \ol{C~f}; \ol{M} \}}  \qquad
\introduceY{m}{\last{D}}
\end{array}
}{\masf{B$_0$}\ \masf{m(\ol{B~x}) \{ return t$_0$; \}} \mbox{ OK} \dashv \qtm{\Phi}{C}}$
}\\[4ex]
\multicolumn{2}{c}{
$\dfrac{
\begin{array}{c}
\masf{\ol{x : B},\,this : C}\ \vdash\ \masf{t$_0$}\ \masf{: E$_0$} \qquad \masf{E$_0$}\, <:\, \masf{B$_0$}\\
\mathit{CT}(\qtm{\Phi}{C}) = \masf{class C extends D \{ \ol{C~f}; \ol{M} \}}  \qquad
\overrideY{m}{\last{D}}{\mol{B} \fct \masf{B$_0$}}
\end{array}
}{\masf{overrides B$_0$}\ \masf{m(\ol{B~x}) \{ return t$_0$; \}} \mbox{ OK} \dashv \qtm{\Phi}{C}}$
}\\[4ex]
\multicolumn{2}{c}{
$\dfrac{
\begin{array}{c}
\masf{\ol{x : B},\,this : C}\ \vdash\ \masf{t$_0$}\ \masf{: E$_0$} \qquad \masf{E$_0$}\, <:\, \masf{B$_0$}\\
\mathit{CT}(\qtm{\Phi}{C}) = \masf{refines class C \{ \ol{C~f}; \ol{M} \}} \qquad
\introduceY{m}{\pred{C}{\Phi}} 
\end{array}
}{\masf{B$_0$}\ \masf{m(\ol{B~x}) \{ return t$_0$; \}} \mbox{ OK} \dashv \qtm{\Phi}{C}}$
}\\[4ex]
\multicolumn{2}{c}{
$\dfrac{
\begin{array}{c}
\masf{\ol{x : B},\,this : C}\ \vdash\ \masf{t$_0$}\ \masf{: E$_0$} \qquad \masf{E$_0$}\, <:\, \masf{B$_0$} \\
\mathit{CT}(\qtm{\Phi}{C}) = \masf{refines class C \{ \ol{C~f}; \ol{M} \}}  \qquad
\overrideY{m}{\pred{C}{\Phi}}{\mol{B} \fct \masf{B$_0$}}
\end{array}
}{\masf{overrides B$_0$}\ \masf{m(\ol{B~x}) \{ return t$_0$; \}} \mbox{ OK} \dashv \qtm{\Phi}{C}}$
}\\[4ex]
\textit{Class typing} & \hfill \fbox{$\masf{L} \mbox{ OK} \dashv \man{\Phi}$} \\[1ex]
\multicolumn{2}{c}{
$\dfrac{
\forall\, \masf{f} \in \mol{f} \, : \, \mathit{introduce}(\masf{f},\last{D}) \qquad \mathit{introduce}(\qtm{\Phi}{C}) \qquad \mol{M} \mbox{ OK} \dashv \qtm{\Phi}{C}
}{\masf{class C extends D \{ \ol{C~f}; \ol{M} \}} \mbox{ OK} \dashv \man{\Phi}}$
}\\[4ex]
\textit{Refinement typing} & \hfill \fbox{$\masf{R} \mbox{ OK} \dashv \man{\Phi}$} \\[1ex]
\multicolumn{2}{c}{
$\dfrac{
\begin{array}{c}
\forall\, \masf{f} \in \mol{f} \, : \, \mathit{introduce}(\masf{f},\pred{C}{\Phi}) \qquad \mathit{refine}(\qtm{\Phi}{C}) \qquad \mol{M} \mbox{ OK} \dashv \qtm{\Phi}{C}  \\ 
\end{array}
}{\masf{refines class C \{ \ol{C~f}; \ol{M} \}} \mbox{ OK} \dashv \man{\Phi}}$
}\end{tabular}
\end{center}
\caption{Well-formedness rules of \ffj.}
\label{fig:type2}
\end{figure*}

\subsubsection{Term Typing Rules.}
A term typing judgment is a triple consisting of a typing context $\man{\Gamma}$, a term \m{t}, and a type \m{C} (see~Figure~\ref{fig:type1}).

Rule \textsc{T-Var} checks whether a free variable is contained in the typing context. Rule \textsc{T-Field} checks whether a field access $\masf{t$_0$.f}$ is well-typed. Specifically, it checks whether $\masf{f}$ is declared in the type of $\masf{t$_0$}$ and whether the type $\masf{f}$ equals the type of the entire term. Rule \textsc{T-Invk} checks whether a method invocation $\masf{t$_0$.m}(\mol{t})$ is well-typed. To this end, it checks whether the arguments $\mol{t}$ of the invocation are subtypes of the types of the formal parameters of $\masf{m}$ and whether the return type of $\masf{m}$ equals the type of the entire term. Rule \textsc{T-New} checks whether an object creation $\masf{new C}(\mol{t})$ is well-typed in that it checks whether the arguments $\mol{t}$ of the instantiation of $\masf{C}$ are subtypes of the types $\mol{D}$ of the fields of $\masf{C}$ and whether $\masf{C}$ equals the type of the entire term. The rules \textsc{T-UCast}, \textsc{T-DCast}, and \textsc{T-SCast} check whether casts are well-typed. In each rule, it is checked whether the type $\masf{C}$ the term $\masf{t$_0$}$ is cast to is a subtype, supertype, or unrelated type of the type of $\masf{t$_0$}$ and whether $\masf{C}$ equals the type of the entire term.\footnote{Rule \textsc{T-SCast} is needed only for the small step semantics of \ffj (and \fj) in order to be able to formulate and prove the type preservation property. \ffj (and \fj) programs whose type derivation contains this rule (i.e., the premise $\mathit{stupid\ warning}$ appears in the derivation) are not further considered (cf.~\cite{IPW01}).}

\subsubsection{Well-Formedness Rules.}
In Figure~\ref{fig:type2}, we show \ffj's well-formedness rules of classes, refinements, and methods. 

The typing judgments of classes and refinements are binary relations between a class or refinement declaration and a feature, written $\masf{L} \mbox{ OK} \dashv \man{\Phi}$ and $\masf{R} \mbox{ OK} \dashv \man{\Phi}$. The rule of classes checks whether all methods are well-formed in the context of the class' qualified type. Moreover, it checks whether none of the fields of the class declaration is introduced multiple times in the combined inheritance and refinement hierarchy and whether there is no feature other than $\man{\Phi}$ that introduces a class $\masf{C}$ (using $\mathit{introduce}$). The well-formedness rule of refinements is analogous, except that the rule checks whether a corresponding class has been introduced before (using $\mathit{refine}$).

The typing judgment of methods is a binary relation between a method declaration and the qualified type that declares the method, written $\masf{M} \mbox{ OK} \dashv \qtm{\Phi}{C}$. There are four different rules for methods (from top to bottom in Figure~\ref{fig:type2}) 
\begin{enumerate}
\item that do not override another method and that are declared by classes,
\item that override another method and that are declared by classes,
\item that do not override another method and that are declared by refinements,
\item that override another method and that are declared by refinements.
\end{enumerate}
All four rules check whether the type $\masf{E$_0$}$ of the method body is a subtype of the declared return type $\masf{B$_0$}$ of the method declaration. For methods that are being introduced, it is checked whether no method with an identical name has been introduced in a superclass (Rule~1) or in a predecessor in the refinement chain (Rule~3). For methods that override other methods, it is checked whether a method with identical name and signature exists in the superclass (Rule~2) or in a predecessor in the refinement chain (Rule~4). 

\subsubsection{Well-Typed \ffj Programs.}
Finally, an \ffj program, consisting of a term, a class table, and a refinement table, is well-typed if
\begin{itemize}
\item the term is well-typed (checked using \ffj's term typing rules),
\item all classes and refinements stored in the class table are well-typed (checked using \ffj's well-formedness rules), and
\item the class and refinement tables are well-formed (ensured by the corresponding sanity conditions).
\end{itemize}

\subsubsection{Type Soundness of \ffj.}
The type system of \ffj is sound. We can prove this using the standard theorems of preservation and progress~\cite{WF94}: 
\begin{thm}[\textit{Preservation}]
If $\man{\Gamma} \vdash \masf{t : C}$ and $\masf{t $\longrightarrow$ t}'$, then $\man{\Gamma} \vdash \masf{t}' \masf{: C}'$ for some $\masf{C}' \masf{<: C}$.
\end{thm}

\begin{thm}[\textit{Progress}]
Suppose $\masf{t}$ is a well-typed term. 
\begin{enumerate}
\item If $\masf{t}$ includes $\masf{new C$_0$(\ol{t}).f$_i$}$ as a subterm, then $\fieldsY{\last{C$_0$}} = \mol{C~f}$ for some $\mol{C}$ and $\mol{f}$.
\item If $\masf{t}$ includes $\masf{new C$_0$(\ol{t}).m(\ol{u})}$ as a subterm, then $\mbodyB{m}{\last{C$_0$}} = (\mol{x}, \masf{t$_0$})$ and $\vert \mol{x} \vert = \vert \mol{u} \vert$ for some $\mol{x}$ and $\masf{t$_0$}$.
\end{enumerate}
\end{thm}
\noindent We provide the proofs of the two theorems in Appendix~\ref{app:ffj_ts}.

\section{Feature-Oriented Product Lines in \ffjpl}
\label{sec:ffjpl}
In this section, our goal is to define a type system for feature-oriented product lines -- a type system that checks whether all valid combinations of features yield well-typed programs. In this scenario, the features in question may be optional or mutually exclusive so that different combinations are possible that form different feature-oriented programs. Since there may be plenty of valid combinations, type checking all of them individually is usually not feasible.

In order to provide a type system for feature-oriented product lines, we need information about which combinations of features are valid, i.e., which features are mandatory, optional, or mutually exclusive, and we need to adapt the subtype and type rules of \ffj to check that there are no combinations/variants that lead to ill-typed terms. The type system guarantees that every program derived from a well-typed product line is a well-typed \ffj program. \ffj together with the type system for checking feature-oriented product lines is henceforth called \ffjpl. 

\subsection{An Overview of Feature-Oriented Product Lines}
\begin{figure}[tbh]
\centering
\includegraphics[scale=0.6,angle=0]{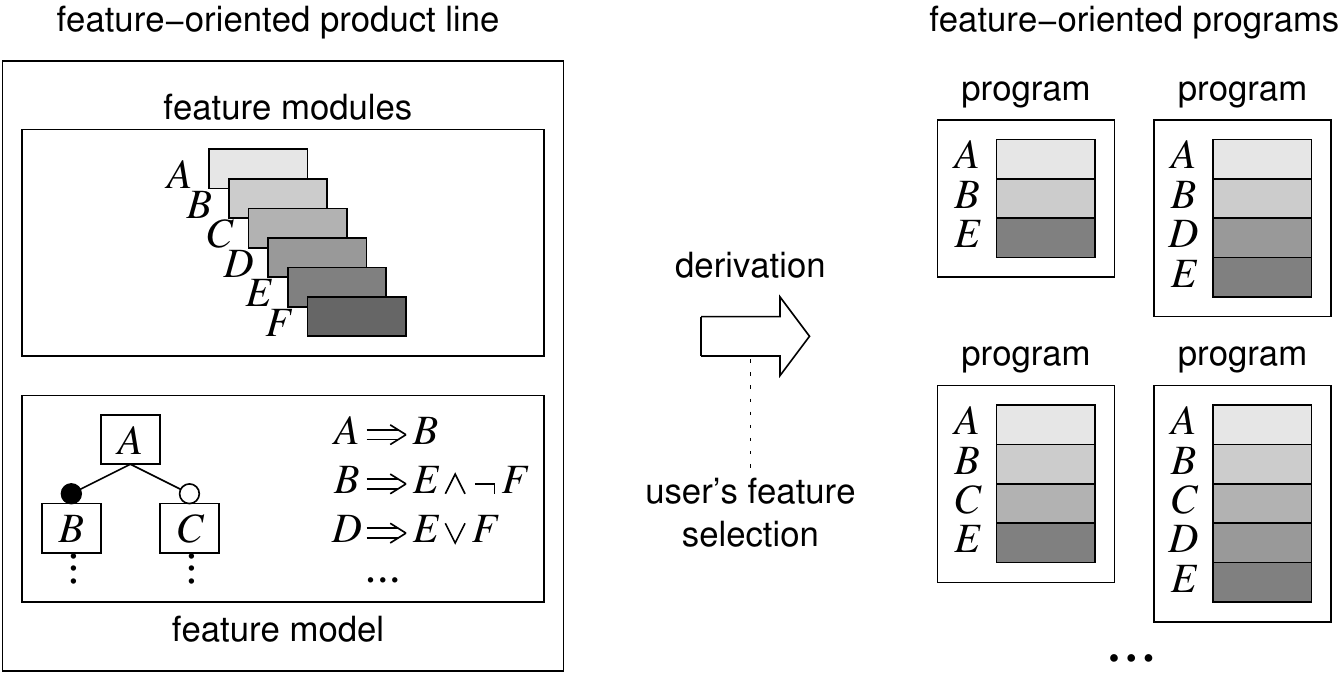}
\caption{The process of deriving programs from a product line.}
\label{fig:derivation}
\end{figure}

A feature-oriented product line is made up of a set of feature modules and a feature model. The feature modules contains the features' implementation and the feature model describes how the feature modules can be combined. In contrast to the feature-oriented programs of Section~\ref{sec:ffj}, typically, some features are optional and some are mutually exclusive (Also other relations such as disjunction, negation, and implication are possible~\cite{Bat05}; they are broken down to mandatory, optional, and mutually exclusive features, as we will explain.). Generally, in a \emph{derivation step}, a user selects a valid subset of features from which, subsequently, a feature-oriented program is derived. In our case, derivation means assembling the corresponding feature modules for a given set of features. In Figure~\ref{fig:derivation}, we illustrate the process of \emph{program derivation}.

Typically, a wide variety of programs can be derived from a product line~\cite{CE00,CN02}. The challenge is to define a type system that guarantees, on the basis of the feature modules and the feature model, that all valid programs are well-typed. Once a program is derived from such a product line, we can be sure that it is well-typed and we can evaluate it using the standard evaluation rules of \ffj (see~Section~\ref{sec:eval}).

\subsection{Managing Variability -- Feature Models}
The aim of developing a product line is to manage the \emph{variability} of a set of programs developed for a particular domain and to facilitate the \emph{reuse} of feature implementations among the programs of the domain. A \emph{feature model} captures the variability by (explicitly or implicitly) defining an ordered set of all features of a product line and their legal feature combinations. A well-defined feature order is essential for field and method lookup (see~Section~\ref{sec:refpl}).

Different approaches to product line engineering use different representations of feature models to define legal feature combinations. The simplest approach is to enumerate all legal feature combinations. In practice, commonly different flavors of tree structures are used, sometimes in combination with additional propositional constraints, to define legal combinations~\cite{CE00,Bat05}, as illustrated in Figure~\ref{fig:derivation}.

For our purpose, the actual representation of legal feature combinations is not relevant. In \ffjpl, we use the feature model only to check whether feature and/or specific program elements are present in certain circumstances. A design decision of \ffjpl is to abstract from the concrete representation of the underlying feature model and rather to provide an interface to the feature model. This has to benefits: (1) we do not need to struggle with all the details of the formalization of feature models, which is well understood by researchers~\cite{Bat05,CP06,TBK+07,DCB09} and outside the scope of this paper, and (2) we are able to support different kinds of feature model representations, e.g., a tree structures, grammars, or propositional formulas~\cite{Bat05}. The interface to the feature model is simply a set of functions and predicates that we use to ask questions like ``may (or may not) feature $\man{A}$ be present together with feature $\man{B}$'' or ``is program element $\masf{m}$ present in every variant in which also feature $\man{A}$ is present'', i.e., ``is program element $\masf{m}$ always \emph{reachable} from feature $\man{A}$''.

\subsection{Challenges of Type Checking}
Let us explain the challenges of type checking by extending our email example, as shown in Figure~\ref{fig:email_ffjpl}. Suppose our basic email client is refined to process incoming text messages (feature \feature{Text}, Lines~1--8). Optionally, it is enabled to process HTML messages, using either Mozilla's rendering engine (feature \feature{Mozilla}, Lines~9--12) or Safari's rendering engine (feature \feature{Safari}, Lines~13--16). To this end, the features \feature{Mozilla} and \feature{Safari} override the method \code{render} of class \code{Display} (Line~11 and~15) in order to invoke the respective rendering engines (field \code{renderer}, Lines~10 and~14) instead of the text printing function (Line~7). 
\begin{figure}[tbh]
\centering
\fheader{Text}
\begin{lstlisting}[firstnumber=1,frame=top|bottom,basicstyle=\sf\scriptsize,
keywordstyle=\sf\scriptsize\bfseries,commentstyle=\sf\scriptsize\it]{ar}
refines class Trans {
  Unit receive(Msg msg) {
    return /* do something... */ new Display().render(msg);
  }
}
class Display {	
  Unit render(Msg msg) { /* display message in text format */ } 
}
\end{lstlisting}
\fheader{Mozilla}
\begin{lstlisting}[frame=top|bottom,basicstyle=\sf\scriptsize,
keywordstyle=\sf\scriptsize\bfseries,commentstyle=\sf\scriptsize\it]{ar}
refines class Display {
  MozillaRenderer renderer;
  overrides Unit render(Msg m) { /* render HTML message using the Mozilla engine */ }
}
\end{lstlisting}
\fheader{Safari}
\begin{lstlisting}[frame=top|bottom,basicstyle=\sf\scriptsize,
keywordstyle=\sf\scriptsize\bfseries,commentstyle=\sf\scriptsize\it]{ar}
refines class Display {
  SafariRenderer renderer;
  overrides Unit render(Msg m) { /* render HTML message using the Safari engine */ }
}
\end{lstlisting}
\vspace{-2ex}
\caption{A feature-oriented email client using Mozilla's and Safari's rendering engines.}
\label{fig:email_ffjpl}
\end{figure}

The first thing to observe is that the features \feature{Mozilla} and \feature{Safari} rely on class \code{Display} and its method \code{render} introduced by feature \feature{Text}. In order to guarantee that every derived program is well-formed, the type system checks whether \code{Display} and \code{render} are \emph{always reachable} from the features \feature{Mozilla} and \feature{Safari}, i.e., whether, in every program variant that contains \feature{Mozilla} and \feature{Safari}, also feature \feature{Text} is present.

The second thing to observe is that the features \feature{Mozilla} and \feature{Safari} both add a field \code{renderer} to \code{Display} (Lines~10 and~14), both of which have different types. In \ffj, a program with both feature modules would not be a well-typed program because the field \code{renderer} is introduced twice. However, Figure~\ref{fig:email_ffjpl} is not intended to represent a single feature-oriented program but a feature-oriented product line; the features \feature{Mozilla} and \feature{Safari} are mutually exclusive, as defined in the product line's feature model (stated earlier), and the type system has to take this fact into account. 

Let us summarize the key challenges of type checking product lines:
\begin{itemize}
\item A global class table contains classes and refinements of all features of a product line, even if some features are optional or mutually exclusive so that they are present only in \emph{some} derived programs. That is, a single class can be introduced by multiple features as long as the features are mutually exclusive. This is also the case for multiple introductions of methods and fields, which may even have different types.
\item The presence of types, fields, and methods depends on the presence of the features that introduce them. A reference from the elements of a feature to a type, a field projection, or a method invocation is valid if the referenced element is always reachable from the referring feature, i.e., in every variant that contains the referring feature.
\item Like references, an extension of a program element, such as a class or method refinement, is valid only if the extended program element is always reachable from the feature that applies the refinement.
\item Refinements of classes and methods do not necessarily form linear refinement chains. There may be alternative refinements of a single class or method that exclude one another, as explained below.
\end{itemize}

\subsection{Collecting Information on Feature Modules}
\label{sec:ctffjpl}
For type checking, the \ffjpl compiler collects various information on the feature modules of the product line. Before the actual type checking is performed, the compiler fills three tables with information: the class table ($\mathit{CT}$), the introduction table ($\mathit{IT}$), and the refinement table ($\mathit{RT}$).

The class table $\mathit{CT}$ of \ffjpl is like the one of \ffj and has to satisfy the same sanity conditions except that (1) there may be multiple declarations of a class (or field or method), as long as they are defined in are mutually exclusive features, and (2) there may be cycles in the inheritance hierarchy, but no cycles for each set of classes which are reachable from any given feature. 

The introduction table $\mathit{IT}$ maps a type to a list $\overline{\man{\Phi}}$ of (mutually exclusive) features that introduce the type. The features returned by $\mathit{IT}$ are listed in the order prescribed by the feature model. In our example of Figure~\ref{fig:email_ffjpl}, a call of $\mathit{IT}(\masf{Display})$ would return a list consisting only of the single feature \feature{Text}. Likewise, the introduction table maps field and method names, in combination with their declaring classes, to features. For example, a call of $\mathit{IT}(\masf{Display}.\masf{renderer})$ would return the list \feature{Mozilla},$\,$\feature{Safari}. The sanity conditions for the introduction table are straightforward:
\begin{itemize}
\item $\mathit{IT}(\masf{C}) = \overline{\man{\Phi}}$ for every type $\masf{C} \in \mathit{dom}(\mathit{CT})$, with $\overline{\man{\Phi}}$ being the features that introduce class $\masf{C}$.
\item $\mathit{IT}(\masf{C.f}) = \overline{\man{\Phi}}$ for every field $\masf{f}$ contained in some class $\masf{C} \in \mathit{dom}(\mathit{CT})$, with $\overline{\man{\Phi}}$ being the features that introduce field $\masf{f}$.
\item $\mathit{IT}(\masf{C.m}) = \overline{\man{\Phi}}$ for every method $\masf{m}$ contained in some class $\masf{C} \in \mathit{dom}(\mathit{CT})$, with $\overline{\man{\Phi}}$ being the features that introduce method $\masf{m}$.
\end{itemize}

Much like in \ffj, in \ffjpl there is a refinement table $\mathit{RT}$. A call of $\mathit{RT}(\masf{C})$ yields a list of all features that either introduce \emph{or} refine class $\masf{C}$, which is different from the introduction table that returns only the features that introduce class $\masf{C}$. As with $\mathit{IT}$, the features returned by $\mathit{RT}$ are listed in the order prescribed by the feature model. The sanity condition for \ffjpl's refinement table is identical to the one of \ffj, namely:
\begin{itemize}
\item $\mathit{RT}(\masf{C}) = \overline{\man{\Phi}}$ for every type $\masf{C} \in \mathit{dom}(\mathit{CT})$, with $\overline{\man{\Phi}}$ being the features that introduce and refine class $\masf{C}$.
\end{itemize}

\subsection{Feature Model Interface}
\label{sec:fm_interface}
As said before, in \ffjpl, we abstract from the concrete representation of the feature model and define instead an interface consisting of proper functions and predicates. There are two kinds of questions we want to ask about the feature model, which we explain next.

First, we would like to know which features are \emph{never} present together, which features are \emph{sometimes} present together, and which features are \emph{always} present together. To this end, we define two predicates, $\mathit{never}$ and $\mathit{sometimes}$, and a function $\mathit{\always}$. Predicate $\mathit{never}(\overline{\man{\Omega}}, \man{\Phi})$ indicates that feature $\man{\Phi}$ is never reachable in the context $\overline{\man{\Omega}}$, i.e., there is no valid program variant in which the features $\overline{\man{\Omega}}$ and feature $\man{\Phi}$ are present together. Predicate $\mathit{sometimes}(\overline{\man{\Omega}}, \man{\Phi})$ indicates that feature $\man{\Phi}$ is sometimes present when the features $\overline{\man{\Omega}}$ are present, i.e., there are variants in which the features $\overline{\man{\Omega}}$ and feature $\man{\Phi}$ are present together and there are variants in which they are not present together. Function $\mathit{\always}(\overline{\man{\Omega}}, \man{\Phi})$ is used to evaluate whether feature $\man{\Phi}$ is always present in the context $\overline{\man{\Omega}}$ (either alone or within a group of alternative features). There are three cases: if feature $\man{\Phi}$ is always present in the context, $\mathit{\always}$ returns the feature again ($\mathit{\always}(\overline{\man{\Omega}},\man{\Phi}) = \man{\Phi}$); if feature $\man{\Phi}$ is not always present, but would be together with a certain group of mutually exclusive features $\overline{\man{\Psi}}$ (i.e., one of the group is always present), $\mathit{\always}$ returns all features of this group ($\mathit{\always}(\overline{\man{\Omega}},\man{\Phi}) = \man{\Phi}, \overline{\man{\Psi}}$). If a feature is not present at all, neither alone nor together with other mutually exclusive features, $\mathit{\always}$ returns the empty list ($\mathit{\always}(\overline{\man{\Omega}},\man{\Phi}) = \bullet$). The above predicates and function provide all information we need to know about the features' relationships. They are used especially for field and method lookup.

Second, we would like to know whether a specific program element is always present when a given set of features is present. This is necessary to ensure that references to program elements are always valid (i.e., not dangling). We need two sources of information for that. First, we need to know all features that introduce the program element in question (determined using the introduction table) and, second, we need to know which combinations of features are legal (determined using the feature model). For the field \code{renderer} of our example, the introduction table would yield the features \feature{Mozilla} and \feature{Safari} and, from the feature model, it follows that \feature{Mozilla} and \feature{Safari} are mutually exclusive, i.e., $\mathit{never}($\feature{Mozilla},$\,$\feature{Safari}$)$. But it can happen that none of the two features is present, which can invalidate a reference to the field. The type system needs to know about this situation.

To this end, we introduce a predicate $\mathit{\validref}$ that expresses that a program element is always reachable from a set of features. For example, $\mathit{\validref}(\overline{\man{\Omega}},\masf{C})$ holds if type \m{C} is always reachable from the context $\overline{\man{\Omega}}$, $\mathit{\validref}(\overline{\man{\Omega}},\masf{C.f})$ holds if field \m{f} of class \m{C} is always reachable from the context $\overline{\man{\Omega}}$, and $\mathit{\validref}(\overline{\man{\Omega}},\masf{C.m})$ holds if method \m{m} of class \m{C} is always reachable from the context $\overline{\man{\Omega}}$. Applying $\mathit{\validref}$ to a list of program elements means that the conjunction of the predicates for every list element is taken. Finally, when we write  $\mathit{\validref}(\overline{\man{\Omega}},\masf{C}) \dashv \overline{\man{\Psi}}$, we mean that program element $\masf{C}$ is always reachable from a context $\overline{\man{\Omega}}$ in a subset $\overline{\man{\Psi}}$ of features of the product line.

In our prototype, we have implemented the above functions and predicates using a SAT solver that reasons about propositional formulas representing constraints on legal feature combinations (see~Section~\ref{sec:impl}), as proposed by Batory~\cite{Bat05} and Czarnecki and Pietroszek~\cite{CP06}.

\subsection{Refinement in \ffjpl}
\label{sec:refpl}
\begin{figure}[tbh]
\centering
\includegraphics[scale=0.6,angle=0]{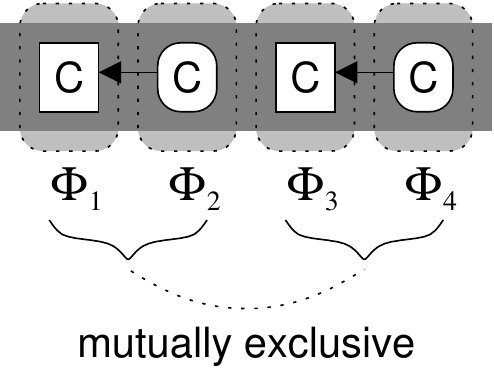}
\caption{Multiple alternative refinements.}
\label{fig:alternative_refinements}
\end{figure}

In Figure~\ref{fig:refpl}, we show the functions $\mathit{last}$ and $\mathit{pred}$ for the navigation along the refinement chain. The two functions are identical to the ones of \ffj (cf.~Figure~\ref{fig:ref}). However, in \ffjpl, there may be alternative declarations of a class and, in the refinement chain, refinement declarations may even precede class declarations, as long as the declaring features are mutually exclusive. Let us illustrate refinement in \ffjpl by means of the example shown in Figure~\ref{fig:alternative_refinements}. Class $\masf{C}$ is introduced in the features $\man{\Phi}_1$ and $\man{\Phi}_3$. Feature $\man{\Phi}_2$ refines class $\masf{C}$ introduced by feature $\man{\Phi}_1$ and feature $\man{\Phi}_4$ refines class $\masf{C}$ introduced by feature $\man{\Phi}_3$. Feature $\man{\Phi}_1$ and $\man{\Phi}_2$ are never present when feature $\man{\Phi}_3$ or $\man{\Phi}_4$ are present and vice versa. A call of $\mathit{RT}(\masf{C})$ would return the list $\man{\Phi}_1, \ldots, \man{\Phi}_4$, a call of $\mathit{last}(\masf{C})$ would return the qualified type $\man{\Phi}_4.\masf{C}$, and a call of $\mathit{pred}(\man{\Phi}_4.\masf{C})$ would return the qualified type $\man{\Phi}_3.\masf{C}$ and so on.
\begin{figure*}
\begin{center}
\begin{tabular}{l l}
\textit{Navigating along the refinement chain} &  \\[2ex]
\multicolumn{2}{c}{
$\dfrac{
\mathit{RT}(\masf{C}) = \overline{\man{\Psi}}
}
{
\last{C} = \man{\Psi}_n.\masf{C}
}$
\qquad
$\dfrac{
\mathit{RT}(\masf{C}) = \overline{\man{\Psi}}, \man{\Phi}, \overline{\man{\Omega}} \qquad \overline{\man{\Psi}} \neq \bullet
}
{
\pred{C}{\Phi} = \man{\Psi}_n.\masf{C} 
}$ 
\qquad
$\dfrac{
\mathit{RT}(\masf{C}) = \man{\Phi}, \overline{\man{\Omega}}
}
{
\pred{C}{\Phi} = \man{Base}.\masf{Object}
}$ 
}
\end{tabular}
\end{center}
\caption{Refinement in \ffjpl.}
\label{fig:refpl}
\end{figure*}

\subsection{Subtyping in \ffjpl}
\begin{figure}[tbh]
\centering
\includegraphics[scale=0.6,angle=0]{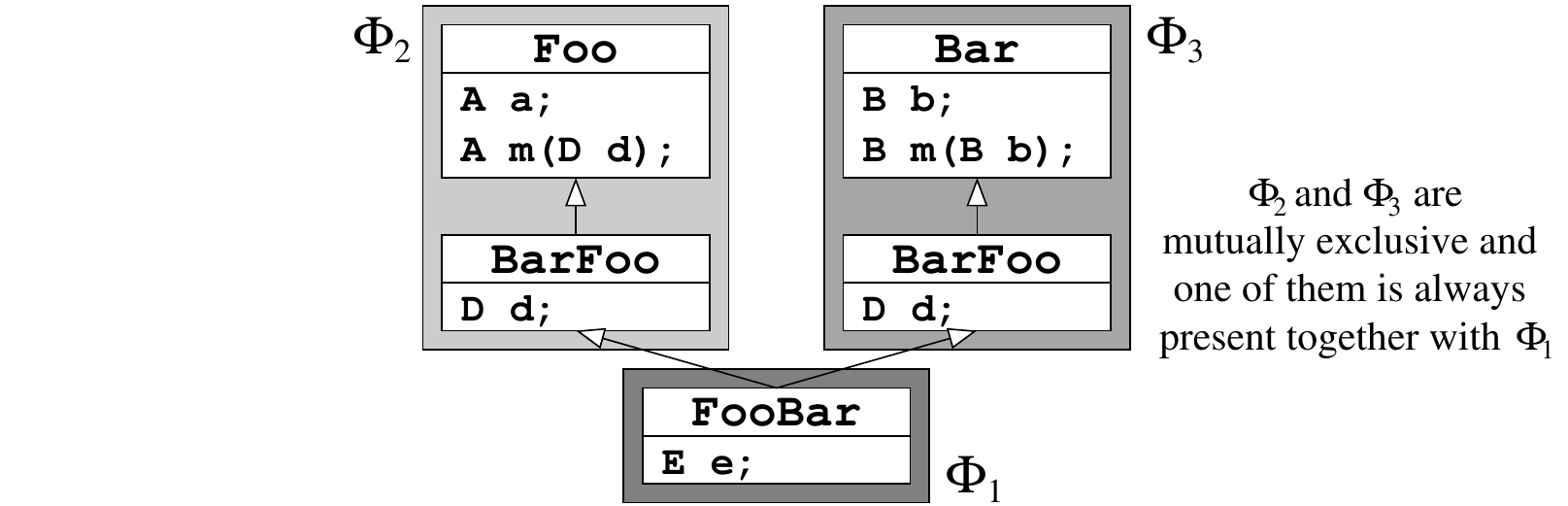}
\caption{Multiple inheritance chains in the presence of alternative features.}
\label{fig:alternative_inheritance}
\end{figure}

The subtype relation is more complicated in \ffjpl than in \ffj. The reason is that a class may have multiple declarations in different features, each declaring possibly different superclasses, as illustrated in Figure~\ref{fig:alternative_inheritance}. That is, when checking whether a class is a subtype of another class, we need to check whether the subtype relation holds in \emph{all} alternative inheritance paths that may be reached from a given context. For example, \code{FooBar} is a subtype of \code{BarFoo} because \code{BarFoo} is a superclass of \code{FooBar} in every program variant (since $\mathit{\always}(\man{\Phi}_1,\man{\Phi}_2)=\man{\Phi}_2,\man{\Phi}_3$); but \code{FooBar} is not a subtype of \code{Foo} and \code{Bar} because, in both cases, a program variant exists in which \code{FooBar} is not a (indirect) subclass of the class in question.

In Figure~\ref{fig:subpl}, we show the subtype relation of \ffjpl. The subtype relation $\masf{C} <:\, \masf{E} \dashv \overline{\man{\Omega}}$ is read as follows: in the context $\overline{\man{\Omega}}$, type $\masf{C}$ is a subtype of type $\masf{E}$, i.e., type $\masf{C}$ is a subtype of type $\masf{E}$ in every variant in which also the features $\overline{\man{\Omega}}$ are present. The first rule in Figure~\ref{fig:subpl} covers reflexivity and terminates the recursion over the inheritance hierarchy. The second rule states that class $\masf{C}$ is a subtype of class $\masf{E}$ if at least one declaration of $\masf{C}$ is always present (tested with $\mathit{\validref}$) and if every of $\masf{C}$'s declarations that may be present together with $\overline{\man{\Omega}}$ (tested with $\mathit{sometimes}$) declares some type $\masf{D}$ as its supertype and $\masf{D}$ is a subtype of $\masf{E}$ in the context $\overline{\man{\Omega}}$. That is, $\masf{E}$ must be a direct or indirect supertype of $\masf{D}$ in \emph{all} variants in which the features $\overline{\man{\Omega}}$ are present. Additionally, supertype $\masf{D}$ must be always reachable from the context ($\overline{\man{\Omega}},\man{\Psi}$). When traversing the inheritance hierarchy, in each step, the context is extended by the feature that introduces the current class in question, e.g., $\overline{\man{\Omega}}$ is extended with $\man{\Psi}$.

Interestingly, the second rule subsumes the two \ffj rules for transitivity and direct superclass declaration because some declarations of $\masf{C}$ may declare $\masf{E}$ directly as its superclass and some declarations may declare another superclass $\masf{D}$ that is, in turn, a subtype of $\masf{E}$, and the rule must be applicable to both cases simultaneously. 
\begin{figure*}
\begin{center}
\begin{tabular}{l l}
\textit{Subtyping} & \hfill \fbox{$\masf{C}\, <:\, \masf{E} \dashv \overline{\man{\Omega}}$} \\[2ex]
\multicolumn{2}{c}{
$\masf{C} <:\, \masf{C} \dashv \overline{\man{\Omega}}$
}\\[4ex]
\multicolumn{2}{c}{
$\dfrac{
\begin{array}{c}
\mathit{\validref}(\overline{\man{\Omega}},\masf{C}) \\[0.5em]
\forall\, \man{\Psi} \in \mathit{IT}(\masf{C}) \, : \, \mathit{sometimes}(\overline{\man{\Omega}},\man{\Psi}) \Rightarrow 
\left(
\begin{array}{c}
\mait{CT}(\qtm{\Psi}{C}) = \masf{class C extends D \{ \ldots\}} \\
\mathit{\validref}((\overline{\man{\Omega}},\man{\Psi}),\masf{D}) \quad \masf{D} <: \masf{E} \dashv \overline{\man{\Omega}},\man{\Psi}
\end{array}
\right)
\end{array}
}
{\masf{C}\, <:\, \masf{E} \dashv \overline{\man{\Omega}}
}$
}\end{tabular}
\end{center}
\caption{Subtyping in \ffjpl.}
\label{fig:subpl}
\end{figure*}

Applied to our example of Figure~\ref{fig:alternative_inheritance}, we have $\mbox{\code{FooBar}} <:\, \mbox{\code{FooBar}} \dashv \man{\Phi}_1$ because of the reflexivity rule. We also have $\mbox{\code{FooBar}} <:\, \mbox{\code{BarFoo}} \dashv \man{\Phi}_1$ because \code{FooBar} is reachable from feature $\man{\Phi}_1$ and every feature that introduces \code{FooBar}, namely $\man{\Phi}_1$, contains a corresponding class declaration that declares \code{BarFoo} as \code{FooBar}'s superclass, and \code{BarFoo} is always reachable from $\man{\Phi}_1$. However, we have $\mbox{\code{FooBar}} \not<:\, \mbox{\code{Foo}} \dashv \man{\Phi}_1$ and $\mbox{\code{FooBar}} \not<:\, \mbox{\code{Bar}} \dashv \man{\Phi}_1$ because \code{FooBar}'s immediate superclass \code{BarFoo} is not always a subtype of \code{Foo} respectively of \code{Bar}.

\subsection{Auxiliary Definitions of \ffjpl}
Extending \ffj toward \ffjpl makes it necessary to add and modify some auxiliary functions. The most complex changes concern the field and method lookup mechanisms.

\subsubsection{Field Lookup.}
\label{sec:fieldspl}
The auxiliary function $\mathit{fields}$ collects the fields of a class including the fields of its superclasses and refinements. Since alternative class or refinement declarations may introduce alternative fields (or the same field with identical or alternative types), $\mathit{fields}$ may return different fields for different feature selections. Since we want to type-check all valid variants, $\mathit{field}$ returns multiple field lists (i.e., a list of lists) that cover all possible feature selections. Each inner list contains field declarations collected in an alternative path of the combined inheritance and refinement hierarchy. 

For legibility, we separate the inner lists using the delimiter `$\fieldsep$'. For example, looking up the fields of class \code{FooBar} in the context of feature $\man{\Phi}_1$ (Figure~\ref{fig:alternative_inheritance}) yields the list $\masf{A~a},\masf{D~d},\masf{E~e}\fieldsep\,\masf{B~b},\masf{D~d},\masf{E~e}$ because the features $\man{\Phi}_2$ and $\man{\Phi}_3$ are mutually exclusive and one of them is present in each variant in which also $\man{\Phi}_1$ is present. For readability, we use the metavariables $\f$ and $\g$ when referring to inner field lists. We abbreviate a list of lists of fields $\f_{1}\fieldsep \ldots\fieldsep \f_{n}$ by $\overline{\f}$. Analogously, $\overline{\overline{\f}}$ is shorthand for $\f_{11}\fieldsep \ldots\fieldsep \f_{n1}\fieldsep \ldots\fieldsep \f_{1m}\fieldsep \ldots\fieldsep \f_{nm}$.

Function $\mathit{fields}$ receives a qualified type $\qtm{\Phi}{C}$ and a context of selected features $\overline{\man{\Omega}}$. If we want all possible field lists, the context is empty. If we want only field lists for a subset of feature selections, e.g., only the fields that can be referenced from a term in a specific feature module, we can use the context to specify one or more features of which we know that they must be selected.

The basic idea of \ffjpl's field lookup is to traverse the combined inheritance and refinement hierarchy much like in \ffj. There are four situations that are handled differently:

\begin{enumerate}
\item The field lookup returns the empty list when it reaches \qt{Base}{Object}.
\item The field lookup ignores all fields that are introduced by features that are never present in a given context.
\item The field lookup collects all fields that are introduced by features that are always present in a given context. References to these fields are always valid.
\item The field lookup collects all fields that are introduced by features that may be present in a given context but that are not always present. In this case, a special marker $@$ is added to the fields in question because we cannot guarantee that a reference to this field is safe in the given context.\footnote{Note that the marker $@$ is generated during type checking, so we do not include it in the syntax of \ffj.} It is up to the type system to decide, based on the marker, whether this situation may provoke an error (e.g., the type system ignores the marker when looking for duplicate fields but reports an error when type checking object creations).
\item A special situation occurs when the field lookup identifies a group of alternative features. In such a group each feature is optional and excludes every other feature of the group and at least one feature of the group is always present in a given context. Once the field lookup identifies a group of alternative features, we split the result list, each list containing the fields of a feature of the group and the fields of the original list.
\end{enumerate}

\begin{figure*}
\begin{center}
\begin{tabular}{l l}
\textit{Field lookup} & \hfill \fbox{$\mathit{fields}(\overline{\man{\Omega}},\qtm{\Phi}{C}) = \overline{\mol{C~f}}$} \\[2ex]
\multicolumn{2}{c}{
\hfill
$\mathit{fields}(\overline{\man{\Omega}},\qtm{\Phi}{Object}) = \bullet$
\hfill (FL-1)
}\\[2ex]
\multicolumn{2}{c}{
\hfill
$\dfrac{
\mathit{never}(\overline{\man{\Omega}},\man{\Phi})
}{
\mathit{fields}(\overline{\man{\Omega}},\qtm{\Phi}{C}) = \mathit{fields}(\overline{\man{\Omega}}, \mathit{pred}(\qtm{\Phi}{C}))
}$
\hfill (FL-2)
}\\[4ex]
\multicolumn{2}{c}{
\hfill
$\dfrac{
\begin{array}{c}
\mathit{sometimes}(\overline{\man{\Omega}}, \man{\Phi}) \qquad \mathit{\always}(\overline{\man{\Omega}}, \man{\Phi}) = \man{\Phi} \\
\mathit{CT}(\qtm{\Phi}{C}) = \masf{class C extends D \{\ } \mol{C~f};\ \mol{M}\ \masf{\}} 
\end{array}
}{
\mathit{fields}(\overline{\man{\Omega}}, \qtm{\Phi}{C}) = \mathit{append}(\mathit{fields}(\overline{\man{\Omega}}, \mathit{last}(\masf{D})), \mol{C~f})
}
$
\hfill (FL-3.1)
}\\[4ex]
\multicolumn{2}{c}{
\hfill
$\dfrac{
\begin{array}{c}
\mathit{sometimes}(\overline{\man{\Omega}}, \man{\Phi}) \qquad \mathit{\always}(\overline{\man{\Omega}}, \man{\Phi}) = \man{\Phi} \\
\mathit{CT}(\qtm{\Phi}{C}) = \masf{refines class C \{\ } \mol{C~f};\ \mol{M}\ \masf{\}}
\end{array}
}{
\mathit{fields}(\overline{\man{\Omega}}, \qtm{\Phi}{C}) = \mathit{append}(\mathit{fields}(\overline{\man{\Omega}}, \mathit{pred}(\qtm{\Phi}{C})), \mol{C~f})
}
$
\hfill (FL-3.2)
}\\[4ex]
\multicolumn{2}{c}{
\hfill
$\dfrac{
\begin{array}{c}
\mathit{sometimes}(\overline{\man{\Omega}}, \man{\Phi}) \qquad \mathit{\always}(\overline{\man{\Omega}}, \man{\Phi}) = \bullet \\
\mathit{CT}(\qtm{\Phi}{C}) = \masf{class C extends D \{\ } \mol{C~f};\ \mol{M}\ \masf{\}}
\end{array}
}{
\mathit{fields}(\overline{\man{\Omega}}, \qtm{\Phi}{C}) = \mathit{append}(\mathit{fields}(\overline{\man{\Omega}}, \mathit{last}(\masf{D})), \overline{\masf{C~f}}@)
}$
\hfill (FL-4.1)
}\\[4ex]
\multicolumn{2}{c}{
\hfill
$\dfrac{
\begin{array}{c}
\mathit{sometimes}(\overline{\man{\Omega}}, \man{\Phi}) \qquad \mathit{\always}(\overline{\man{\Omega}}, \man{\Phi}) = \bullet \\
\mathit{CT}(\qtm{\Phi}{C}) = \masf{refines class C \{\ } \mol{C~f};\ \mol{M}\ \masf{\}}
\end{array}
}{
\mathit{fields}(\overline{\man{\Omega}}, \qtm{\Phi}{C}) = \mathit{append}(\mathit{fields}(\overline{\man{\Omega}}, \mathit{pred}(\qtm{\Phi}{C})), \overline{\masf{C~f}}@)
}$
\hfill (FL-4.2)
}\\[4ex]
\multicolumn{2}{c}{
\hfill
$
\dfrac{
\mathit{sometimes}(\overline{\man{\Omega}}, \man{\Phi}) \qquad \mathit{\always}(\overline{\man{\Omega}}, \man{\Phi})=\overline{\man{\Psi}} 
}{
\mathit{fields}(\overline{\man{\Omega}}, \qtm{\Phi}{C}) = \mathit{fields}((\overline{\man{\Omega}},\man{\Psi}_1), \qtm{\Phi}{C})\fieldsep\ldots\fieldsep\mathit{fields}((\overline{\man{\Omega}},\man{\Psi}_n), \qtm{\Phi}{C})
}
$
\hfill (FL-5)
}
\end{tabular}
\end{center}
\caption{Field lookup in \ffjpl.}
\label{fig:fieldspl}
\end{figure*}

In order to distinguish the different cases, we use the predicates and functions defined in Section~\ref{sec:fm_interface} (especially $\mathit{never}$, $\mathit{sometimes}$, and $\mathit{\always}$). The definition of function $\mathit{fields}$, shown in Figure~\ref{fig:fieldspl}, follows the intuition described above:  Once \qt{Base}{Object} is reached, the recursion terminates (FL-1). When a feature is never reachable in the given context, $\mathit{fields}$ ignores this feature and resumes with the previous one (FL-2). When a feature is mandatory (i.e., always present in a given context), the fields in question are added to each alternative result list, which were created in Rule FL-5 (FL-3.1 and FL-3.2).\footnote{Function $\mathit{append}$ adds to each inner list of a list of field lists a given field. Its implementation is straightforward and omitted for brevity.} When a feature is optional, the fields in question, annotated with the marker $@$, are added to each alternative result list (FL-4.1 and FL-4.2). When a feature is part of an alternative group of features, we cannot immediately decide how to proceed. We split the result list in multiple lists (by means of multiple recursive invocations of $\mathit{fields}$), in which we add one of the alternative features to each context passed to an invocation of $\mathit{fields}$ (FL-5).

\subsubsection{Method Type Lookup.}
\begin{figure*}
\begin{center}
\begin{tabular}{l l}
\textit{Method type lookup} & \hfill \fbox{$\mathit{mtype}(\overline{\man{\Omega}},\masf{m},\qtm{\Phi}{C}) = \overline{\overline{\masf{B}} \fct \masf{B$_0$}}$} \\[2ex]
\multicolumn{2}{c}{
\hfill
$\mathit{mtype}(\overline{\man{\Omega}},\masf{m},\qtm{Base}{Object}) = \bullet$
\hfill (ML-1)
}\\[2ex]
\multicolumn{2}{c}{
\hfill
$\dfrac{
\begin{array}{c}
\masf{B$_0$ m(\ol{B~x}) \{ \ldots \}} \in \mol{M} \qquad \mathit{sometimes}(\overline{\man{\Omega}},\man{\Phi})\\
\mathit{CT}(\qtm{\Phi}{C}) = \masf{class C extends D \{\ } \mol{C~f};\ \mol{M}\ \masf{\}} 
\end{array}
}
{\mathit{mtype}(\overline{\man{\Omega}},\masf{m},\qtm{\Phi}{C}) = \mathit{mtype}(\overline{\man{\Omega}},\masf{m},\mathit{pred}(\qtm{\Phi}{C})), \mathit{mtype}(\overline{\man{\Omega}},\masf{m},\mathit{last}(\masf{D})), \mol{B} \fct \masf{B$_0$}}$
\hfill (ML-2)
}\\[4ex]
\multicolumn{2}{c}{
\hfill
$\dfrac{
\begin{array}{c}
\masf{B$_0$ m(\ol{B~x}) \{ \ldots \}} \in \mol{M} \qquad \mathit{sometimes}(\overline{\man{\Omega}},\man{\Phi}) \\
\mathit{CT}(\qtm{\Phi}{C}) = \masf{refines class C \{\ } \mol{C~f};\ \mol{M}\ \masf{\}}
\end{array}
}
{\mathit{mtype}(\overline{\man{\Omega}},\masf{m},\qtm{\Phi}{C}) = \mathit{mtype}(\overline{\man{\Omega}},\masf{m},\mathit{pred}(\qtm{\Phi}{C})), \mol{B} \fct \masf{B$_0$}}$
\hfill (ML-3)
}\\[4ex]
\multicolumn{2}{c}{
\hfill
$\dfrac{
\begin{array}{c}
(\masf{m} \mbox{ is not defined in } \mol{M} \ \vee \ \mathit{never}(\overline{\man{\Omega}},\man{\Phi})) \\
\mathit{CT}(\qtm{\Phi}{C}) = \masf{class C extends D \{\ } \mol{C~f};\ \mol{M}\ \masf{\}} 
\end{array}
}
{\mathit{mtype}(\overline{\man{\Omega}},\masf{m},\qtm{\Phi}{C}) = \mathit{mtype}(\overline{\man{\Omega}},\masf{m},\mathit{pred}(\qtm{\Phi}{C})), \mathit{mtype}(\overline{\man{\Omega}},\masf{m},\mathit{last}(\masf{D}))
}$
\hfill (ML-4)
}\\[4ex]
\multicolumn{2}{c}{
\hfill
$\dfrac{
\begin{array}{c}
 (\masf{m} \mbox{ is not defined in } \mol{M} \ \vee \ \mathit{never}(\overline{\man{\Omega}},\man{\Phi})) \\
\mathit{CT}(\qtm{\Phi}{C}) = \masf{refines class C \{\ } \mol{C~f};\ \mol{M}\ \masf{\}}
\end{array}
}
{\mathit{mtype}(\overline{\man{\Omega}},\masf{m},\qtm{\Phi}{C}) = \mathit{mtype}(\overline{\man{\Omega}},\masf{m},\mathit{pred}(\qtm{\Phi}{C}))
}$
\hfill (ML-5)
}
\end{tabular}
\end{center}
\caption{Method Lookup in \ffjpl.}
\label{fig:methodspl}
\end{figure*}
Like in field lookup, in method lookup, we have to take alternative definitions of methods into account. But the lookup mechanism is simpler than in $\mathit{fields}$ because the order of signatures found in the combined inheritance and refinement hierarchy is irrelevant for type checking. Hence, function $\mathit{mtype}$ yields a simple list $\overline{\mol{B} \fct \masf{B$_0$}}$ of signatures for a given method name $\masf{m}$. For example, calling $\mathit{mtype}(\man{\Phi}_1,\masf{m},\man{\Phi}_1.\masf{C})$ in the context of Figure~\ref{fig:alternative_inheritance} yields the list $\masf{D} \fct \masf{A}, \masf{B} \fct \masf{B}$. 

In Figure~\ref{fig:methodspl}, we show the definition of function $\mathit{mtype}$. For $\man{Base}.\masf{Object}$, the empty list is returned (ML-1). If a class that is sometimes reachable introduces a method in question (ML-2), its signature is added to the result list and all possible predecessors in the refinement chain (using $\mathit{pred}$) and all possible subclasses are searched (using $\mathit{last}$). Likewise, if a refinement that is sometimes reachable introduces a method with the name searched (ML-3), its signature is added to the result list and all possible predecessors in the refinement chain are searched (using $\mathit{pred}$). If a class or refinement does not declare a corresponding method (ML-4 and ML-5) or the a class is never reachable, the search proceeds with the possible superclasses or predecessors. 

The current definition of function $\mathit{mtype}$ returns possibly many duplicate signatures. A straightforward optimization would be to remove duplicates before using the result list, which we omitted for simplicity. 

\subsubsection{Valid Introduction, Refinement, and Overriding.}
\begin{figure*}
\begin{center}
\begin{tabular}{l l}
\multicolumn{1}{l}{\textit{Valid class introduction}} \hfill & \hfill \fbox{$\mathit{introduce}(\overline{\man{\Omega}},\qtm{\Phi}{C})$} \\[2ex]
\multicolumn{2}{c}{
$\dfrac{
\nexists\, \man{\Psi}\, :\, \left(
\begin{array}{c}
\mathit{CT}(\qtm{\Psi}{C}) = \masf{class C extends D \{\ } \mol{C~f};\ \mol{M}\ \masf{\}} \\
\man{\Psi} \ne \man{\Phi} \qquad \mathit{sometimes}(\overline{\man{\Omega}},\man{\Psi})
\end{array}
\right)
}
{\mathit{introduce}(\overline{\man{\Omega}},\qtm{\Phi}{C})}$
}\\[4ex]
\textit{Valid field introduction} & \hfill \fbox{$\mathit{introduce}(\overline{\man{\Omega}},\masf{f},\qtm{\Phi}{C})$} \\[2ex]
\multicolumn{2}{c}{
$\dfrac{
\forall\, \mol{E~h} \in \mathit{fields}(\overline{\man{\Omega}},\qtm{\Phi}{C})\, :\, \masf{f} \notin \mol{h}
}
{\mathit{introduce}(\overline{\man{\Omega}},\masf{f},\qtm{\Phi}{C})}$
}\\[4ex]
\textit{Valid method introduction} & \hfill \fbox{$\mathit{introduce}(\overline{\man{\Omega}},\masf{m},\qtm{\Phi}{C})$} \\[2ex]
\multicolumn{2}{c}{
$\dfrac{
\mathit{mtype}(\overline{\man{\Omega}},\masf{m},\qtm{\Phi}{C}) = \bullet
}
{\mathit{introduce}(\overline{\man{\Omega}},\masf{m},\qtm{\Phi}{C})}$
}\\[4ex]
\multicolumn{1}{l}{\textit{Valid class refinement}} \hfill & \hfill \fbox{$\mathit{refine}(\overline{\man{\Omega}},\qtm{\Phi}{C})$} \\[2ex]
\multicolumn{2}{c}{
$\dfrac{
\mathit{RT}(\masf{C}) = \overline{\man{\Psi}}, \man{\Phi}, \overline{\man{\Pi}} \qquad \mathit{\validref}(\overline{\man{\Omega}},\masf{C}) \dashv \overline{\man{\Psi}}
}
{\mathit{refine}(\overline{\man{\Omega}},\qtm{\Phi}{C})}$
}\\[4ex]
\textit{Valid method overriding} & \hfill \fbox{$\mathit{override}(\overline{\man{\Omega}},\masf{m},\qtm{\Phi}{C},\mol{C} \fct \masf{C$_0$})$} \\[2ex]
\multicolumn{2}{c}{
$\dfrac{
\begin{array}{c}
\mathit{RT}(\masf{C}) = \overline{\man{\Psi}}, \man{\Phi}, \overline{\man{\Pi}} \qquad \mathit{\validref}(\overline{\man{\Omega}},\masf{C.m}) \dashv \overline{\man{\Psi}},\man{\Phi} \\
\forall\, \mol{B} \fct \masf{B$_0$} \in \mathit{mtype}(\overline{\man{\Omega}},\masf{m},\qtm{\Phi}{C}) \, : \, \mol{C} = \mol{B} \ \wedge \ \masf{C$_0$} = \masf{B$_0$}
\end{array}
}
{\mathit{override}(\overline{\man{\Omega}},\masf{m},\qtm{\Phi}{C},\mol{C} \fct \masf{C$_0$})}$
}
\end{tabular}
\end{center}
\caption{Valid introduction, refinement, and overriding in \ffjpl.}
\label{fig:validspl}
\end{figure*}
In Figure~\ref{fig:validspl}, we show predicates for checking the validity of introduction, refinement, and overriding in \ffjpl. 
Predicate $\mathit{introduce}$ indicates whether a class with the qualified type \qt{\Phi}{C} has not been introduced by any other feature $\man{\Psi}$ that may be present in the context $\overline{\man{\Omega}}$. Likewise, $\mathit{introduce}$ holds if a method $\masf{m}$ or a field $\masf{f}$ has not been introduced by a qualified type $\qtm{\Phi}{C}$ (including possible predecessors and superclasses) that may be present in the given context $\overline{\man{\Omega}}$. To this end, it checks either whether $\mathit{mtype}$ yields the empty list or whether $\masf{f}$ is not contained in every inner list returned by $\mathit{fields}$.

For a given refinement, predicate $\mathit{refine}$ indicates whether a proper class, which is always reachable in the given context, has been declared previously in the refinement chain. We write $\mathit{\validref}(\overline{\man{\Omega}},\masf{C}) \dashv \overline{\man{\Psi}}$ in order to state that a declaration of class $\masf{C}$ has been introduced in the set $\overline{\man{\Psi}}$ of features, which is only a subset of the features of the product line, namely the features that precede the feature that introduces class \m{C}. Predicate $\mathit{override}$ indicates whether a declaration of method $\masf{m}$ has been introduced (and is always reachable) in some feature introduced by before the feature that refines \m{m} and whether every possible declaration of $\masf{m}$ in any predecessor of a $\qtm{\Phi}{C}$ has the same signature.

\subsection{Type Relation of \ffjpl}
\label{sec:ffjpl_tr}
\begin{figure*}
\begin{center}
\begin{tabular}{l l}
\textit{Term typing} & \hfill \fbox{$\man{\Gamma}\ \vdash\ \masf{t : } \mol{C} \dashv \man{\Phi}$} \\[1ex]
\multicolumn{2}{c}{
\hfill 
$\dfrac{\masf{x : C} \in \man{\Gamma}}{\man{\Gamma}\ \vdash\ \masf{x : } \masf{C} \dashv \man{\Phi}}$ \hfill ({\sc T-Var}$_{\mathit{PL}}$)
}\\[4ex]
\multicolumn{2}{c}{
\hfill 
$\dfrac{
\begin{array}{c}
\forall\, \masf{E} \in \mol{E} \, : \, \mathit{\validref}(\man{\Phi},\masf{E.f}) \\
\man{\Gamma}\ \vdash\ \masf{t$_0$}\ \masf{: } \mol{E} \dashv \man{\Phi} \qquad \mathit{fields}(\man{\Phi},\overline{\mathit{last}(\masf{E})}) = \overline{\overline{\f,\masf{C~f},\g}}
\end{array}
}
{
\man{\Gamma}\ \vdash\ \masf{t$_0$} \masf{.f}\ \masf{: } \masf{C}_{11}, \ldots, \masf{C}_{n1}, \ldots, \masf{C}_{1m}, \ldots, \masf{C}_{nm}  \dashv \man{\Phi}
}$ \hfill ({\sc T-Field}$_{\mathit{PL}}$)
}\\[4ex]
\multicolumn{2}{c}{
\hfill $\dfrac{
\begin{array}{c}
\forall\, \masf{E} \in \mol{E} \, : \, \mathit{\validref}(\man{\Phi},\masf{E.m}) \qquad
\forall\, \mol{C} \in \overline{\mol{C}},\, \forall\, \mol{D} \in \overline{\mol{D}} \in \overline{\overline{\mol{D}}}\, :\, \overline{\masf{C} <: \masf{D}} \dashv \man{\Phi} \\
\man{\Gamma}\ \vdash\ \masf{t$_0$ : } \mol{E} \dashv \man{\Phi} \qquad \man{\Gamma}\ \vdash\ \overline{\masf{t : } \mol{C}} \dashv \man{\Phi} \qquad \mathit{mtype}(\man{\Phi},\masf{m},\overline{\mathit{last}(\masf{E})}) = \overline{\overline{\mol{D} \fct \masf{B}}}
\end{array}
}
{
\man{\Gamma}\ \vdash\ \masf{t$_0$}\masf{.m(\ol{t}) : } \masf{B}_{11}, \ldots, \masf{B}_{n1}, \ldots, \masf{B}_{1m}, \ldots, \masf{B}_{nm} \dashv \man{\Phi}
}$ \hfill ({\sc T-Invk}$_{\mathit{PL}}$)
}\\[4ex]
\multicolumn{2}{c}{
\hfill 
$\dfrac{
\begin{array}{c}
\mathit{\validref}(\man{\Phi},\masf{C}) \qquad
\forall\, \mol{D~g} \in \overline{\f}\,,\, \forall\, \mol{C} \in \overline{\mol{C}}\, :\, \overline{\masf{C} <: \masf{D}} \dashv \man{\Phi} \\
\man{\Gamma}\ \vdash\ \overline{\masf{t : } \mol{C}} \dashv \man{\Phi} \qquad \mathit{fields}(\man{\Phi},\mathit{last}(\masf{C})) = \overline{\f} \qquad @ \notin \overline{\f}
\end{array}
}
{
\man{\Gamma}\ \vdash\ \masf{new C(\ol{t}) : } \masf{C} \dashv \man{\Phi}
}$ \hfill ({\sc T-New}$_{\mathit{PL}}$)
}\\[4ex]
\multicolumn{2}{c}{
\hfill $\dfrac{
\begin{array}{c}
\mathit{\validref}(\man{\Phi},\masf{C})\\
\man{\Gamma}\ \vdash\ \masf{t}_0\ \masf{: } \mol{E} \dashv \man{\Phi} \qquad \forall\, \masf{E} \in \mol{E}\, : \, ( \masf{E}\, <:\, \masf{C}  \dashv \man{\Phi}\ \vee\ \masf{C}\, <:\, \masf{E}  \dashv \man{\Phi})
\end{array}
}
{
\man{\Gamma}\ \vdash\ \masf{(C)t}_0\ \masf{: } \masf{C} \dashv \man{\Phi}
}$ \hfill ({\sc T-UDCast}$_{\mathit{PL}}$)
}\\[4ex]
\multicolumn{2}{c}{
\hfill $\dfrac{
\begin{array}{c}
\mathit{\validref}(\man{\Phi},\masf{C}) \qquad \mathit{stupid~warning} \\
\man{\Gamma}\ \vdash\ \masf{t}_0\ \masf{: } \mol{E} \dashv \man{\Phi} \qquad \exists\, \masf{E} \in \mol{E}\, : \, (\masf{C}\, \not <:\, \masf{E}  \dashv \man{\Phi}\ \wedge\ \masf{E}\, \not <:\, \masf{C}  \dashv \man{\Phi})
\end{array}
}
{
\man{\Gamma}\ \vdash\ \masf{(C)t}_0\ \masf{: } \masf{C} \dashv \man{\Phi}
}$ \hfill ({\sc T-SCast}$_{\mathit{PL}}$)
}
\end{tabular}
\end{center}
\caption{Term typing in \ffjpl.}
\label{fig:typepl1}
\end{figure*}

\begin{figure*}
\begin{center}
\begin{tabular}{l l}
\textit{Method typing} & \hfill \fbox{$\masf{M} \mbox{ OK} \dashv \qtm{\Phi}{C}$} \\[1ex]
\multicolumn{2}{c}{
$\dfrac{
\begin{array}{c}
\masf{\ol{x : B},\,this : C} \vdash \masf{t$_0$}\ \masf{: } \mol{E} \dashv \man{\Phi} \qquad \forall\, \masf{E} \in \mol{E}\, :\, \masf{E}\, <:\, \masf{B$_0$} \dashv \man{\Phi} \\
\mathit{\validref}(\man{\Phi},\mol{B}) \qquad
\mathit{introduce}(\man{\Phi},\masf{m},\mathit{last}(\masf{D})) \\
\mathit{CT}(\qtm{\Phi}{C}) = \masf{class C extends D \{ \ol{C~f}; \ol{M} \}}
\end{array}
}{\masf{B$_0$}\ \masf{m(\ol{B~x}) \{ return t$_0$; \}} \mbox{ OK} \dashv \qtm{\Phi}{C}}$
}\\[4ex]
\multicolumn{2}{c}{
$\dfrac{
\begin{array}{c}
\masf{\ol{x : B},\,this : C} \vdash \masf{t$_0$}\ \masf{: } \mol{E} \dashv \man{\Phi} \qquad \forall\, \masf{E} \in \mol{E}\, :\, \masf{E}\, <:\, \masf{B$_0$} \dashv \man{\Phi} \\
\mathit{\validref}(\man{\Phi},\mol{B}) \qquad
\mathit{override}(\man{\Phi},\masf{m},\mathit{last}(\masf{D}),\mol{B} \fct \masf{B$_0$})\\
\mathit{CT}(\qtm{\Phi}{C}) = \masf{class C extends D \{ \ol{C~f}; \ol{M} \}}
\end{array}
}{\masf{overrides B$_0$}\ \masf{m(\ol{B~x}) \{ return t$_0$; \}} \mbox{ OK} \dashv \qtm{\Phi}{C}}$
}\\[4ex]
\multicolumn{2}{c}{
$\dfrac{
\begin{array}{c}
\masf{\ol{x : B},\,this : C} \vdash \masf{t$_0$}\ \masf{: } \mol{E} \dashv \man{\Phi} \qquad \forall\, \masf{E} \in \mol{E}\, :\, \masf{E}\, <:\, \masf{B$_0$} \dashv \man{\Phi} \\
\mathit{\validref}(\man{\Phi},\mol{B}) \qquad
\mathit{introduce}(\man{\Phi},\masf{m},\mathit{pred}(\qtm{\Phi}{C})) \\
\mathit{CT}(\qtm{\Phi}{C}) = \masf{refines class C \{ \ol{C~f}; \ol{M} \}}
\end{array}
}{\masf{B$_0$}\ \masf{m(\ol{B~x}) \{ return t$_0$; \}} \mbox{ OK} \dashv \qtm{\Phi}{C}}$
}\\[4ex]
\multicolumn{2}{c}{
$\dfrac{
\begin{array}{c}
\masf{\ol{x : B},\,this : C} \vdash \masf{t$_0$}\ \masf{: } \mol{E} \dashv \man{\Phi} \qquad \forall\, \masf{E} \in \mol{E}\, :\, \masf{E}\, <:\, \masf{B$_0$} \dashv \man{\Phi} \\
\mathit{\validref}(\man{\Phi},\mol{B}) \qquad
\mathit{override}(\man{\Phi},\masf{m},\mathit{pred}(\qtm{\Phi}{C}),\mol{B} \fct \masf{B$_0$})\\
\mathit{CT}(\qtm{\Phi}{C}) = \masf{refines class C \{ \ol{C~f}; \ol{M} \}}
\end{array}
}{\masf{overrides B$_0$}\ \masf{m(\ol{B~x}) \{ return t$_0$; \}} \mbox{ OK} \dashv \qtm{\Phi}{C}}$
}\\[4ex]
\textit{Class typing} & \hfill \fbox{$\masf{L} \mbox{ OK} \dashv \man{\Phi}$} \\[1ex]
\multicolumn{2}{c}{
$\dfrac{
\begin{array}{c}
\mathit{\validref}(\man{\Phi},\masf{D}) \qquad \mathit{\validref}(\man{\Phi},\mol{C}) \\
\forall\, \masf{f} \in \mol{f} \, : \, \mathit{introduce}(\man{\Phi},\masf{f},\mathit{last}(\masf{D})) \qquad \mathit{introduce}(\man{\Phi},\qtm{\Phi}{C}) \qquad \mol{M} \mbox{ OK} \dashv \qtm{\Phi}{C}
\end{array}
}{\masf{class C extends D \{ \ol{C~f}; \ol{M} \}} \mbox{ OK} \dashv \man{\Phi}}$
}\\[4ex]
\textit{Refinement typing} & \hfill \fbox{$\masf{R} \mbox{ OK} \dashv \man{\Phi}$} \\[1ex]
\multicolumn{2}{c}{
$\dfrac{
\begin{array}{c}
\mathit{\validref}(\man{\Phi},\mol{C}) \\
\forall\, \masf{f} \in \mol{f} \, : \, \mathit{introduce}(\man{\Phi},\masf{f},\mathit{pred}(\qtm{\Phi}{C})) \qquad \mathit{refine}(\man{\Phi},\qtm{\Phi}{C}) \qquad \mol{M} \mbox{ OK} \dashv \qtm{\Phi}{C}
\end{array}
}{\masf{refines class C \{ \ol{C~f}; \ol{M} \}} \mbox{ OK} \dashv \man{\Phi}}$
}\\[4ex]
\end{tabular}
\end{center}
\caption{Well-formedness rules of \ffjpl.}
\label{fig:typepl2}
\end{figure*}
The type relation of \ffjpl consists of type rules for terms and well-formedness rules for classes, refinements, and methods, shown in Figure~\ref{fig:typepl1} and Figure~\ref{fig:typepl2}.

\subsubsection{Term Typing Rules.}
\label{sec:ffjpl_tt}
A term typing judgment in \ffjpl is a quadruple, consisting of a typing context $\man{\Gamma}$, a term \m{t}, a list of types $\mol{C}$, and a feature $\man{\Phi}$ that contains the term (see~Figure~\ref{fig:typepl1}). A term can have multiple types in a product line because there may be multiple declarations of classes, fields, and methods. The list $\mol{C}$ contains all possible types a term can have.

Rule \textsc{T-Var}$_{\mathit{PL}}$ is standard and does not refer to the feature model. It yields a list consisting only of the type of the variable in question.

Rule \textsc{T-Field}$_{\mathit{PL}}$ checks whether a field access $\masf{t$_0$.f}$ is well-typed in every possible variant in which also $\man{\Phi}$ is present. Based on the possible types $\mol{E}$ of the term $\masf{t$_0$}$ the field $\masf{f}$ is accessed from, the rule checks whether $\masf{f}$ is always reachable from $\man{\Phi}$ (using $\mathit{\validref}$). Note that this is a key mechanism of \ffjpl's type system. It ensures that a field, being accessed, is definitely present in every valid program variant in which the field access occurs -- without generating all these variants. Furthermore, all possible fields of all possible types $\mol{E}$ are assembled in a nested list $\overline{\overline{\f,\masf{C~f},\g}}$ in which $\masf{C~f}$ denotes a declaration of the field \m{f}; the call of
$\mathit{fields}(\man{\Phi},\overline{\mathit{last}(\masf{E})})$ is shorthand for $\mathit{fields}(\man{\Phi},\mathit{last}(\masf{E}_1))\ \ldots\ \mathit{fields}(\man{\Phi},\mathit{last}(\masf{E}_n))$, in which the individual result lists are concatenated. Finally, the list of all possible types $\masf{C}_{11}, \ldots, \masf{C}_{n1}, \ldots, \masf{C}_{1m}, \ldots, \masf{C}_{nm}$ of field \m{f} becomes the list of types of the overall field access. Note that the result list may contain duplicates, which could be eliminated for optimization purposes.

Rule \textsc{T-Invk}$_{\mathit{PL}}$ checks whether a method invocation $\masf{t$_0$.m}(\mol{t})$ is well-typed in every possible variant in which also $\man{\Phi}$ is present. Based on the possible types $\mol{E}$ of the term $\masf{t$_0$}$ the method $\masf{m}$ is invoked on, the rule checks whether $\masf{m}$ is always reachable from $\man{\Phi}$ (using $\mathit{\validref}$). As with field access, this check is essential. It ensures that in generated programs only methods are invoked that are also present. Furthermore, all possible signatures of $\masf{m}$ of all possible types $\mol{E}$ are assembled in the nested list $\overline{\overline{\mol{D} \fct \masf{B}}}$ and it is checked that all possible lists $\mol{C}$ of argument types of the method invocation are subtypes of all possible lists $\mol{D}$ of parameter types of the method (this implies that the lengths of the two lists must be equal). A method invocation has multiple types assembled in a list that contains all result types of method $\masf{m}$ determined by $\mathit{mtype}$. As with field access, duplicates should be eliminated for optimization purposes.

Rule \textsc{T-New}$_{\mathit{PL}}$ checks whether an object creation $\masf{new C}(\mol{t})$ is well-typed in every possible variant in which also $\man{\Phi}$ is present. Specifically, it checks whether there is a declaration of class \m{C} always reachable from $\man{\Phi}$. Furthermore, all possible field combinations of $\masf{C}$ are assembled in the nested list $\overline{\f}$, and it is checked whether all possible combinations of argument types passed to the object creation are subtypes of the types of all possible field combinations (this implies that the number of arguments types must equal the number of field types). The fields of the result list must not be annotated with the marker $@$ since optional fields may not be present in every variant and references may become invalid (see field lookup).\footnote{The treatment of $@$ is semiformal but simplifies the rule.} An object creation has only a single type $\masf{C}$.

Rules \textsc{T-UDCast}$_{\mathit{PL}}$ and \textsc{T-SCast}$_{\mathit{PL}}$ check whether casts are well-typed in every possible variant in which also $\man{\Phi}$ is present. This is done by checking whether the type $\masf{C}$ the term $\masf{t$_0$}$ is cast to is always reachable from $\man{\Phi}$ and whether this type is a subtype, supertype, or unrelated type of all possible types $\mol{E}$ the term $\masf{t$_0$}$ can have. We have only a single rule \textsc{T-UDCast}$_{\mathit{PL}}$ for up- and downcasts because the list $\mol{E}$ of possible types may contain super- and subtypes of $\masf{C}$ simultaneously. If there is a type in the list which leads to a stupid case, we flag a $\mathit{stupid\ warning}$. A cast yields a list containing only a single type $\masf{C}$.

\subsubsection{Well-Formedness Rules.}
\label{sec:ffjpl_wf}
In Figure~\ref{fig:typepl2}, we show the well-formedness rules of classes, refinements, and methods. 

Like in \ffj, the typing judgment of classes and refinements is a binary relation between a class or refinement declaration and a feature. The rule of classes checks whether all methods are well-formed in the context of the class' qualified type. Moreover, it checks whether the class declaration is unique in the scope of the enclosing feature $\man{\Phi}$, i.e., whether no other feature, that may be present together with feature $\man{\Phi}$, introduces a class with an identical name (using $\mathit{introduce}$). Furthermore, it checks whether the superclass and all field types are always reachable from $\man{\Phi}$ (using $\mathit{\validref}$). Finally, it checks whether none of the fields of the class declaration have been introduced before (using $\mathit{introduce}$). The well-formedness rule of refinements is analogous, except that the rule checks that there is \emph{at least one} class declaration reachable that is refined and that has been introduced before the refinement (using $\mathit{refine}$).

The typing judgment of methods is a binary relation between a method declaration and the qualified type that declares the method. Like in \ffj, there are four different rules for methods (from top to bottom in Figure~\ref{fig:typepl2})
\begin{enumerate}
\item that do not override another method and that are declared by classes,
\item that override another method and that are declared by classes,
\item that do not override another method and that are declared by refinements,
\item that override another method and that are declared by refinements.
\end{enumerate}
All four rules check whether all possible types $\mol{E}$ of the method body are subtypes of the declared return type $\masf{B$_0$}$ of the method and whether the argument types $\mol{B}$ are always reachable from the enclosing feature $\man{\Phi}$ (using $\mathit{\validref}$).

For methods that are introduced, it is checked, using $\mathit{introduce}$, whether no method with identical name has been introduced in any possible superclass (Rule~1) or in any possible predecessor in the refinement chain (Rule~3). For methods that override other methods, it is checked, using $\mathit{override}$, whether a method with identical name and signature exists in any possible superclass (Rule~2) or in any possible predecessor in the refinement chain (Rule~4). 

\subsubsection{Well-Typed \ffjpl Product Lines.}
An \ffjpl product line, consisting of a term, a class table, an introduction table, and a refinement table, is well-typed if
\begin{itemize}
\item the term is well-typed (checked using \ffjpl's term typing rules),
\item all classes and refinements stored in the class table are well-formed (checked using \ffjpl's well-formedness rules), and
\item the class, introduction, and refinement tables are well-formed (ensured by the corresponding sanity conditions).
\end{itemize}

\subsection{Type Safety of \ffjpl}
Type checking in \ffjpl is based on information contained in the class table, introduction table, refinement table, and feature model. The first three are filled by the compiler that has parsed the code base of the product line. The feature model is supplied directly by the user (or tool). The compiler determines which class and refinement declarations belong to which features. The classes and refinements of the class table are checked using their well-formedness rules which, in turn, use the well-formedness rules for methods and the term typing rules for method bodies. Several rules use the introduction and refinement tables in order to map types, fields, and methods to features and the feature model to navigate along refinement chains and to check the presence of program elements.

What does type safety mean in the context of a product line? The product line itself is never evaluated; rather, different programs are derived that are then evaluated. Hence, the property we are interested in is that \emph{all} programs that can be derived from a well-typed product line are in turn well-typed. Furthermore, we would like to be sure that \emph{all} \ffjpl product lines, from which only well-typed \ffj programs can be derived, are well-typed. We formulate the two properties as the two theorems \textit{Correctness of} \ffjpl and \textit{Completeness of} \ffjpl.

\newpage

\subsubsection{Correctness}
\begin{thm}[\textit{Correctness of} \ffjpl]
\label{thm:ffjpl_correct}
\normalfont Given a well-typed \ffjpl product line $\mathit{pl}$ (including with a well-typed term $\masf{t}$, well-formed class, introduction, and refinement tables $\mathit{CT}$, $\mathit{IT}$, and $\mathit{RT}$, and a feature model $\mathit{FM}$), every program that can be derived with a valid feature selection $\mathit{fs}$ is a well-typed \ffj program (cf.~Figure~\ref{fig:derivation}).
\begin{equation*}
\dfrac{
\mathit{pl} = (\masf{t}, \mathit{CT},\mathit{IT},\mathit{RT},\mathit{FM}) \qquad \mathit{pl} \mbox{\normalfont~is well-typed } \qquad \mathit{fs} \mbox{\normalfont~is valid in } \mathit{FM}
}
{
\mathit{derive}(\mathit{pl},\mathit{fs}) \mbox{\normalfont~is well-typed}
}
\end{equation*}
\end{thm}

Function $\mathit{derive}$ collects the feature modules from a product line according to a user's selection $\mathit{fs}$, i.e., non-selected feature modules are removed from the derived program. After this derivation step, the class table contains only classes and refinements stemming from the selected feature modules. We define a \emph{valid feature selection} to be a list of features whose combination does not contradict the constraints implied by the feature model.

The proof idea is to show that the type derivation tree of an \ffjpl product line is a superimposition of multiple so-called \emph{type derivation slices}. As usual, the type derivation proceeds from the root (i.e., an initial type rule that checks the term and all classes and refinements of the class table) to the leaves (type rules that do not have a premise) of the type derivation tree. Each time a term has multiple types, e.g., a method has different alternative return types, which is caused by multiple mutually exclusive method declarations, the type derivation splits into multiple branches. With \emph{branch} we refer only to positions in which the type derivation tree is split into multiple subtrees in order to type check multiple mutually exclusive term definitions. Each subtree from the root of the type derivation tree along the branches toward a leaf is a type derivation slice. Each slice corresponds to the type derivation of a feature-oriented program.

Let us illustrate the concept of a type derivation slice by a simplified example. Suppose the application of an arbitrary type rule to a term $\masf{t}$ somewhere in the type derivation. Term $\masf{t}$ has multiple types $\mol{C}$ due to different alternative definitions of $\masf{t}$'s subterms. For simplicity, we assume here that $\masf{t}$ has only a single subterm $\masf{t$_0$}$, like in the case of a field access ($\masf{t} = \masf{t}_0.\masf{f}$), in which the overall term $\masf{t}$ has multiple types depending on $\masf{t}_0$'s and $\masf{f}$'s types; the rule can be easily extended to multiple subterms by adding a predicate per subterm. The type rule ensures the well-typedness of all possible variants of $\masf{t}$ on the basis of the variants of $\masf{t}$'s subterm $\masf{t$_0$}$. Furthermore, the type rule checks whether a predicate $\mathit{predicate}$ (e.g., $\masf{C <: D}$) holds for each variant of the subterm with its possible types $\mol{E}$, written $\mathit{predicate}(\masf{t$_0$ : E$_i$})$. The possible types $\mol{C}$ of the overall term follow in some way from the possible types $\mol{E}$ of its subterm. Predicate $\mathit{\validref}$ is used to check whether all referenced elements and types are present in all valid variants, including different combinations of optional features. For the general case, this can be written as follows:
\begin{equation*}
\qquad \qquad \dfrac{
\begin{array}{c}
\mathit{predicate}(\masf{t$_0$ : E$_1$}) \quad \mathit{predicate}(\masf{t$_0$ : E$_2$}) \quad \ldots \quad \mathit{predicate}(\masf{t$_0$ : E$_n$})\\
\masf{t$_0$} : \mol{E} \qquad \mathit{always}(\ldots)
\end{array}
}
{
\man{\Gamma}\ \vdash\ \masf{t} : \mol{C} \dashv \man{\Phi}
} (\mbox{\sc T-*}_{\mathit{PL}})
\end{equation*}

The different uses of $\mathit{predicate}$ in the premise of an \ffjpl type rule correspond to the branches in the type derivation that denote alternative definitions of subterms. Hence, the premise of the \ffjpl type rule is the conjunction of the different premises that cover the different alternative definitions of the subterms of a term.

The proof strategy is as follows. Assuming that the \ffjpl type system ensures that each slice is a valid \ffj type derivation (see~Lemma~\ref{lem:a} in Appendix~\ref{sec:corr}) and that each valid feature selection corresponds to a single slice (since alternative features have been removed; see~Lemma~\ref{lem:b} in Appendix~\ref{sec:corr}), each feature-oriented program that corresponds to a valid feature selection is guaranteed to be well-typed. Note that multiple valid feature selections may correspond to the same slice because of the presence of optional features. It follows that, for every valid feature selection, we derive a well-formed \ffj program -- since its type derivation is valid -- whose evaluation satisfies the properties of progress and preservation (see~Appendix~\ref{app:ffj_ts}). In Appendix~\ref{app:ffjpl_ts}, we describe the proof of Theorem~\ref{thm:ffjpl_correct} in more detail.

\subsubsection{Completeness}

\begin{thm}[\textit{Completeness of} \ffjpl]
\label{thm:ffjpl_complete}
Given an \ffjpl product line $\mathit{pl}$ (including a well-typed term $\masf{t}$, well-formed class, introduction, and refinement tables $\mathit{CT}$, $\mathit{IT}$, and $\mathit{RT}$, and a feature model $\mathit{FM}$), and given that \emph{all} valid feature selections $\mathit{fs}$ yield well-typed \ffj programs, according to Theorem~\ref{thm:ffjpl_correct}, $\mathit{pl}$ is a well-typed product line according to the rules of \ffjpl.
\begin{equation*}
\dfrac{
\mathit{pl} = (\masf{t},\mathit{CT},\mathit{IT},\mathit{RT},\mathit{FM}) \quad \forall\,\mathit{fs}\,:\,( \mathit{fs} \mbox{\normalfont~is valid in } \mathit{FM} \Rightarrow \mathit{derive}(\mathit{pl},\mathit{fs}) \mbox{\normalfont~is well-typed} )
}
{
\mathit{pl} \mbox{\normalfont~is well-typed }
}
\end{equation*}
\end{thm}

The proof idea is to examine three basic cases and to generalize subsequently: (1) $\mathit{pl}$ has only mandatory features; (2) $\mathit{pl}$ has only mandatory features except a single optional feature; (3) $\mathit{pl}$ has only mandatory features except two mutually exclusive features. All other cases can be formulated as combinations of these three basic cases. To this end, we divide the possible relations between features into three disjoint sets: (1) a feature is reachable from another feature in \emph{all} variants, (2) a feature is reachable from another feature in \emph{some}, but \emph{not in all}, variants, (3) two features are mutually exclusive. From these three possible relations, we can prove the three basic cases in isolation and, subsequently, construct a general case that can be phrased as a combination of the three basic cases. The description of the general case and the reduction finish the proof of Theorem~\ref{thm:ffjpl_complete}. In Appendix~\ref{app:ffjpl_ts}, we describe the proof of Theorem~\ref{thm:ffjpl_complete} in detail.

\section{Implementation \& Discussion}
\label{sec:impl}
\begin{figure}[tbh]
\centering
\includegraphics[scale=0.5,angle=0]{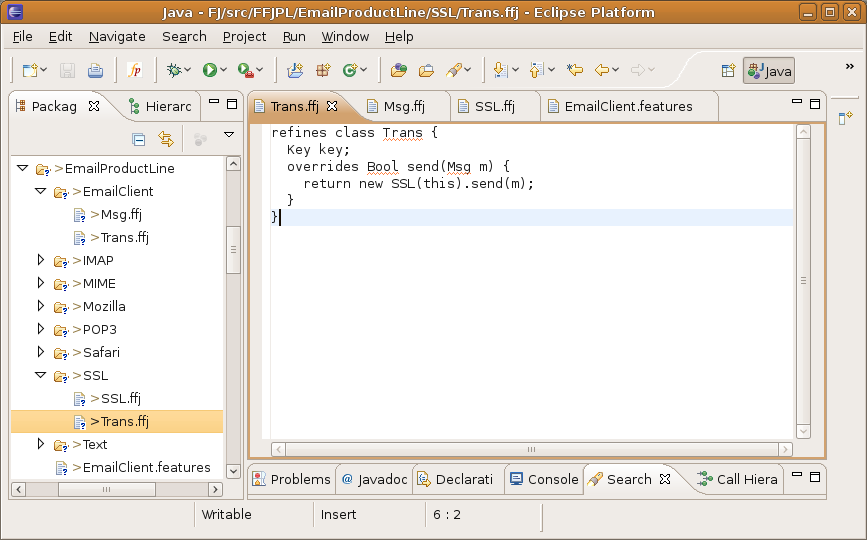}
\caption{Snapshot of the test environment of the Haskell implementation.}
\label{fig:ide}
\end{figure}

We have implemented \ffj and \ffjpl in Haskell, including the program evaluation and type checking of product lines. The \ffjpl compiler expects a set of feature modules and a feature model both of which, together, represent the product line. A feature module is represented by a directory. The files found inside a feature module's directory are assigned to / belong to the enclosing feature. The \ffjpl compiler stores this information for type checking. Each file may contain multiple classes and class refinements. In Figure~\ref{fig:ide}, we show a snapshot of our test environment, which is based on Eclipse and a Haskell plugin\footnote{\url{http://eclipsefp.sourceforge.net/haskell/}}. We use Eclipse to interpret or compile our \ffj and \ffjpl type systems and interpreters. Specifically, the figure shows the directory structure of our email system. The file \code{EmailClient.features} contains the user's feature selection and the feature model of the product line.

The feature model of a product line is represented by a propositional formula, following the approach of Batory~\cite{Bat05} and Czarnecki and Pietroszek~\cite{CP06}. Propositional formulas are an effective way of representing the relationships between features (e.g., of specifying which feature implies the presence and absence of other features and of machine checking whether a feature selection is valid). For example, we have implemented predicate $\mathit{sometimes}$ as follows:
\begin{equation*}
\mathit{sometimes}(\mathit{FM}, \overline{\man{\Omega}}, \man{\Psi}) = \mathit{satisfiable} (\mathit{FM} \wedge \man{\Omega}_1 \wedge \ldots \wedge \man{\Omega}_n \wedge \man{\Psi})
\end{equation*}
The feature model is an propositional formula; feature are variables; and $\mathit{satisfiable}$ is a satisfiability solver. Likewise, we have implemented predicate $\mathit{always}$ on the basis of logical reasoning on propositional formulas:
\begin{equation*}
\mathit{\always}(\mathit{FM}, \overline{\man{\Omega}}, \man{\Psi}) = \neg (\mathit{satisfiable} ( \neg (\mathit{FM} \Rightarrow ((\man{\Omega}_1 \wedge \ldots \wedge \man{\Omega}_n) \Rightarrow \man{\Psi}))))
\end{equation*}
For a more detailed explanation of how propositional formulas relate to feature models and feature selections, we refer the interest to the work of Batory~\cite{Bat05}.

In Figure~\ref{fig:email_fm}, we show the textual specification of the feature model of our email system, which can be passed directly to the \ffjpl compiler.
\begin{figure}[tbh]
\centering
\begin{lstlisting}[firstnumber=1,frame=top|bottom,basicstyle=\sf\scriptsize,
keywordstyle=\sf\scriptsize\bfseries,commentstyle=\sf\scriptsize\it]{}
$\mbox{\bfseries features:}$
  EmailClient IMAP POP3 MIME SSL Text Mozilla Safari

$\mbox{\bfseries model:}$
  EmailClient $\mbox{\bfseries implies}$ (IMAP or POP3);
  IMAP $\mbox{\bfseries implies}$ EmailClient;
  POP3 $\mbox{\bfseries implies}$ EmailClient;
  MIME $\mbox{\bfseries implies}$ EmailClient;
  SSL $\mbox{\bfseries implies}$ EmailClient;
  Text $\mbox{\bfseries implies}$ (IMAP or POP3);
  Mozilla $\mbox{\bfseries implies}$ (IMAP or POP3);
  Safari $\mbox{\bfseries implies}$ (IMAP or POP3);
  Mozilla $\mbox{\bfseries implies}$ (not Safari);
  Safari $\mbox{\bfseries implies}$ (not Mozilla);
\end{lstlisting}
\caption{Feature model of an email client product line.}
\label{fig:email_fm}
\end{figure}

The first section ($\mbox{\bfseries features:}$) of the file representing the feature model defines an ordered set of names of the features of the product line and the second section ($\mbox{\bfseries model:}$) defines constraints on the features' presence in the derived programs. In our example, each email client supports either the protocols IMAP, POP3, or both. Furthermore, every feature requires the presence of the base feature \feature{EmailClient}. Feature \feature{Text} requires either the presence of \feature{IMAP} or \feature{POP3} or both -- the same for \feature{Mozilla} and \feature{Safari}. Finally, feature \feature{Mozilla} requires the absence of feature \feature{Safari} and vice versa.

On the basis of the feature modules and the feature model, \ffjpl's type system checks the entire product line and identifies valid program variants that still contain type errors. A SAT solver is used to check whether elements are never, sometimes, or always reachable. If an error is found, the product line is rejected as ill-formed. If not, a feature-oriented program guaranteed to be well-formed can derived on the basis of a user's feature selection. This program can be evaluated using the standard evaluation rules of \ffj, which we have also implemented in Haskell.

In contrast to previous work on type checking feature-oriented product lines~\cite{TBK+07,DCB09}, our type system provides detailed error messages. This is possible due to the fine-grained checks at the level of individual term typing and well-formedness rules. For example, if a field access succeeds only in some program variants, this fact can be reported to the user and the error message can point to the erroneous field access. Previously proposed type systems compose all code of all features of a product line and extract a single propositional formula, which is checked for satisfiability. If the formula is not satisfiable (i.e., a type error has occurred), it is not possible to identify the location that has caused the error (at least not without further information). See Section~\ref{sec:related_work}, for a detailed discussion of related approaches.

We made several tests and experiments with our Haskell implementation. However, real-world tests were not feasible because of two reasons. First, in previous work it has been already demonstrated that feature-oriented product lines require proper type systems and that type checking entire real-world product lines is feasible and useful~\cite{TBK+07}. Second, like \fj, \ffj is a core language into which all Java programs can be compiled and which, by its relative simplicity, is suited for the formal definition and proof of language properties -- in our case, a type system and its correctness and completeness. But, a core language is never suited for the development of real-world programs. This is why our examples and test programs are of similar size and complexity as the \fj examples of Pierce~\cite{Pie02}. Type checking our test programs required acceptable amounts of time (in the order of magnitude of milliseconds per product line). We do not claim to be able to handle full-sized feature-oriented product lines by hand-coding them in \ffjpl. Rather, this would require an expansion of the type system to full Java (including support for features as provided by AHEAD~\cite{BSR04} or FeatureHouse~\cite{AKL09fh}) -- an enticing goal, but one for the future (especially, as Java's informal language specification~\cite{GJS+05} has 688 pages). Our work lays a foundation for implementing type systems in that it provides evidence that core feature-oriented mechanisms are type sound and type systems of feature-oriented product lines can be implemented correctly and completely.

Still, we would like to make some predictions on the scalability of our approach. The novelty of our type system is that it incorporates alternative features and, consequently, alternative definitions of classes, fields, and methods. This leads to a type derivation tree with possibly multiple branches denoting alternative term types. Hence, performing a type derivation of product line with many alternative features may consume a significant amount of computation time and memory. It seems that this overhead is the price for allowing alternative implementation of program parts. 

Nevertheless, our approach minimizes the overhead caused by alternative features compared to the naive approach. In the naive approach, all possible programs are derived and type checked subsequently. In our approach, we type check the entire code base of the product line and branch the type derivation only at terms that really have multiple, alternative types, and not at the level of entire program variants, as done in the naive approach. 
Our experience with feature-oriented product lines shows that, usually, there are not many alternative features in a product line, but mostly optional features~\cite{LB01,AB05,TBK+07,KAB07,RSS+08fame,ALS08tse,AKL09fh,AJT+09,AKG+09sc,RKS+09btw,SKR09btw}. For example, in the Berkeley DB product line (JE edition; 80\,000 lines of code) there are 99 feature modules, but only two pairs of them alternative~\cite{AKL09fh,KAB07}; in the Graph Product Line there are 26 feature modules, of which only three pairs are alternative~\cite{LB01,AKL09fh}. A further observation is that most alternative features that we encountered do not alter types. That is, there are multiple definitions of fields and methods but with equal types. For example, GPL and Berkeley DB contain alternative definitions of a few methods but only with identical signatures. Type checking these product lines with our approach, the type derivation would have almost no branches. In the naive approach, still many program variants exist due to optional features. Hence, our approach is preferable. For example, in a product line with $n$ features and $c*n$ variants (with $c$ being a constant), in our approach, the type system would have to check $n$ feature modules (with some few branches in the type derivation and solving few simple SAT problems; see below) and, in the naive approach, the type system would have to check, at least, $2*n$ feature modules but, commonly, $2*n*m$ with $m < n$. For product lines with a higher degree of variability, e.g., with $n^2$ or even $2^n$ variants the benefit of our approach becomes even more significant. We believe that this benefit can make a difference in real world product line engineering.

A further point is that almost all typing and well-formedness rules contain calls to the built-in SAT solver. This results in possibly many invocations of the SAT solver at type checking time. Determining the satisfiability of a propositional formula is in general an $\mathcal{NP}$-complete problem. However, it has been shown that the structures of propositional formulas occurring in software product lines are simple enough to scale satisfiability solving to thousands of features~\cite{MWC09sat}. Furthermore, in our experiments, we have observed that many calls to the SAT solver are redundant, which is easy to see when thinking about type checking feature-oriented product lines where the presence of single types or members is checked in many type rules. We have implemented a caching mechanism to decrease the number of calls to the SAT solver to a minimum. 

Finally, the implementation in Haskell helped us a lot with the evaluation of the correctness of our type rules. It can serve other researchers to reproduce and evaluate our work and to experiment with further (feature-oriented) language mechanisms. The implementations of \ffj and \ffjpl, along with test programs, can be downloaded from the Web.\footnote{\url{http://www.fosd.de/ffj}}

\section{Related Work}
\label{sec:related_work}
We divide our discussions of related work into two parts: the implementation, formal models, and type systems (1) of feature-oriented programs and (2) of feature oriented product lines.

\subsection{Feature-Oriented Programs}
\ffj has been inspired by several feature-oriented languages and tools, most notably AHEAD/Jak~\cite{BSR04}, FeatureC++~\cite{ALR+05b}, FeatureHouse~\cite{AKL09fh}, and Prehofer's feature-oriented Java extension~\cite{Pre97}. Their key aim is to separate the implementation of software artifacts, e.g., classes and methods, from the definition of features. That is, classes and refinements are not annotated or declared to belong to a feature. There is no statement in the program text that defines explicitly a connection between code and features. Instead, the mapping of software artifacts to features is established via so-called containment hierarchies, which are basically directories containing software artifacts. The advantage of this approach is that a feature's implementation can include, beside classes in the form of Java files, also other supporting documents, e.g., documentation in the form of HTML files, grammar specifications in the form of JavaCC files, or build scripts and deployment descriptors in the form of XML files~\cite{BSR04}. To this end, feature composition merges not only classes with their refinements but also other artifacts, such as HTML or XML files, with their respective refinements~\cite{ADT07,AKL09fh}.


Another class of programming languages that provide mechanisms for the definition and extension of classes and class hierarchies includes, e.g., \emph{ContextL}~\cite{HCN08}, \emph{Scala}~\cite{OZ05}, and \emph{Classbox/J}~\cite{BDN05}.
The difference to feature-oriented languages is that they provide explicit language constructs for aggregating the classes that belong to a feature, e.g., family classes, classboxes, or layers. This implies that non-code software artifacts cannot be included in a feature~\cite{ALS08tse}. However, \ffj still models a subset of these languages, in particular, class refinement.

Similarly, related work on a formalization of the key concepts underlying feature-oriented programming has not disassociated the concept of a feature from the level of code. Especially, calculi for mixins~\cite{FKF98,BPS99,ALZ03,KT04}, traits~\cite{LS08}, family polymorphism and virtual classes~\cite{ISV05,EOC06,Hut06,CDN+07}, path-dependent types~\cite{OZ05,OCR+03}, open classes~\cite{CML+06}, dependent classes~\cite{GMO07}, and nested inheritance~\cite{NCM04} either support only the refinement of single classes or expect the classes that form a semantically coherent unit (i.e., that belong to a feature) to be located in a physical module that is defined in the host programming language. For example, a virtual class is by definition an inner class of the enclosing object, and a classbox is a package that aggregates a set of related classes. Thus, \ffj differs from previous approaches in that it relies on contextual information that has been collected by the compiler, e.g., the features' composition order or the mapping of code to features.

A different line of research aims at the language-independent reasoning about features~\cite{BSR04,LBL06,AKL09fh,KAT09}. The calculus gDeep is most closely related to \ffj since it provides a type system for feature-oriented languages that is language-independent~\cite{AH07tr}. The idea is that the recursive process of merging software artifacts, when composing hierarchically structured features, is very similar for different host languages, e.g., for Java, C\#, and XML. The calculus describes formally how feature composition is performed and what type constraints have to be satisfied. In contrast, \ffj does not aspire to be language-independent, although the key concepts can certainly be used with different languages. The advantage of \ffj is that its type system can be used to check whether terms of the host language (Java or FJ) violate the principles of feature orientation, e.g., whether methods refer to classes that have been added by other features. Due to its language independence, gDeep does not have enough information to perform such checks.

\subsection{Feature-Oriented Product Lines}

Our work on type checking feature-oriented product lines was motivated by the work of Thaker et al.~\cite{TBK+07}. They suggested the development of a type system for feature-oriented product lines that does not check all individual programs but the individual feature implementations. They have implemented an (incomplete) type system and, in a number of case studies on real product lines, they found numerous hidden errors using their type rules. Nevertheless, the implementation of their type system is ad-hoc in the sense that it is described only informally, and they do not provide a correctness and completeness proof. Our type system has been inspired by their work and we were able to provide a formalization and a proof of type safety.

In a parallel line of work, Delaware et al.\ have developed a formal model of a feature-oriented language, called \emph{Lightweight Feature Java} (\emph{LFJ}), and a type system for feature-oriented product lines~\cite{DCB09}. Their work was also influenced by the practical work of Thaker et al. So, it is not surprising that it is closest to ours. However, there are numerous differences. First, their formal model of a feature-oriented language is based on \emph{Lightweight Java} (\emph{LJ})~\cite{SSP07} and not on \emph{Featherweight Java} (\emph{FJ}). While LJ is more expressive, it is also more complex. We decided for the simpler variant FJ, omitting, e.g., constructors and mutable state. Second, Delaware et al.\ do not model feature-oriented mechanisms, such as class or method refinements, directly in the semantics and type rules of the language. Instead, they introduce a transformation step in which LFJ code is ``compiled down'' to LJ code, i.e., they flatten refinement chains to single classes. Proceeding likewise, we would have to generate first an \fj program from an \ffj product line and type check the \fj program (that consists of some or all possible features of the product line) subsequently. We refrained from such a transformation step in order to model the semantics of feature-oriented mechanisms directly in terms of dedicated field and method lookup mechanisms as well as special well-formed rules for method and class refinements. 

Lagorio et al.\ have shown that a flattening semantics and a direct semantics are equivalent~\cite{LSZ09}. An advantage of a ``direct'' semantics is that it allows a type checking and error reporting at a finer grain. In LFJ, all feature modules are composed and a single propositional formula is generated and tested for satisfiability; if the formula is not satisfiable, it is difficult to identify precisely the point of failure. In \ffjpl, the individual type rules consult the feature model and can point directly to the point of failure. 

A further advantage of our approach is that it leaves open \emph{when} feature composition is performed. Currently, feature composition is modeled in \ffj/\ffjpl as a static process done before compilation but, with our approach, it becomes possible to model dynamic feature composition at run time~\cite{RSS+08,Ost02} by making the class and feature tables and the feature model dynamic, i.e., allowing them to change during a computation. With LFJ this is not possible. Hutchins has shown that feature composition can be performed by an interpreter and partial evaluation can be used to pre-evaluate the parts of a composition that are static~\cite{Hut08}. However, Delaware et al.\ have developed a machine-checked model of their type system formalized with the theorem prover Coq~\cite{BP04}. Our proof is hand-written, but we have a Haskell implementation of the \ffj and \ffjpl calculi that we have tested thoroughly. 

Even previously to the work of Thaker et al., Czarnecki et al.\ presented an automatic verification procedure for ensuring that no ill-structured UML model template instances will be generated from a valid feature selection~\cite{CP06}. That is, they type check product lines that consist not of Java programs but of UML models. They use OCL (object constraint language) constraints to express and implement a type system for model composition. In this sense, their aim is very similar to that of \ffjpl, but limited to model artifacts -- although they have proposed to generalize their work to programming languages.

K{\"a}stner et al.\ have implemented a tool, called CIDE, that allows a developer to decompose a software system into features via annotations~\cite{KAK08}. In contrast to other feature-oriented languages and tools, the link between code and features is established via annotations. If a user selects a set of features, all code that is annotated with features (using background colors) that are not present in the selection is removed. K{\"a}stner et al.\ have developed a formal calculus and a set of type rules that ensure that only well-typed programs can be generated from a valid feature selection~\cite{KA08}. For example, if a method declaration is removed, the remaining code must not contain calls to this method. CIDE's type rules are related to the type rules of \ffjpl but, so far, mutually exclusive features are not supported in CIDE. In some sense, \ffjpl and CIDE represent two sides of the same coin: the former aims at the composition of feature modules, the latter at the annotation of feature-related code.

\section{Conclusion}
A feature-oriented product line imposes severe challenges on type checking. The naive approach of checking all individual programs of a product line is not feasible because of the combinatorial explosion of program variants. Hence, the only practical option is to check the entire code base of a product line, including all features, and, based on the information of which feature combinations are valid, to ensure that it is not possible to derive a valid program variant that contains type errors.

We have developed such a type system based on a formal model of a feature-oriented Java-like language, called Feature Featherweight Java (FFJ). A distinguishing property of our work is that we have modeled the semantics and type rules for core feature-oriented mechanisms directly, without compiling feature-oriented code down to a lower-level representation such as object-oriented Java code. The direct semantics allows us to reason about core feature-oriented mechanisms in terms of themselves and not of generated lower-level code. A further advantage is the fine-grained error reporting and that the time of feature composition may vary between compile time and run time.

We have demonstrated and proved that, based on a valid feature selection, our type system ensures that every program of a feature-oriented product line is well-formed and that our type system is complete. Our implementation of \ffj, including the type system for product lines, indicates the feasibility of our approach and can serve as a testbed for experimenting with further feature-oriented mechanisms.

\subsection*{Acknowledgment}
This work is being funded in part by the German Research Foundation (DFG), project number AP 206/2-1.

\begin{appendix}

\section{Type Soundness Proof of \ffj}
\label{app:ffj_ts}
Before giving the main proof, we state and proof some required lemmas.
\begin{lem}
\label{lem:1}
If $\mtypeY{m}{\mathit{last}(\masf{D})} = \mol{C} \fct \masf{C}_0$, then $\mtypeY{m}{\mathit{last}(\masf{C})} = \mol{C} \fct \masf{C}_0$ for all $\masf{C <: D}$.
\end{lem}
\begin{proof}
Straightforward induction on the derivation of $\masf{C <: D}$. There are two cases: First, if method \m{m} is not defined in the declaration or in any refinement of class \m{C}, then $\mtypeY{m}{\mathit{last}(\masf{C})}$ should be the same as $\mtypeY{m}{\mathit{last}(\masf{E})}$ where $\mathit{CT}(\qtm{\Phi}{C}) = \masf{class C extends E \{ \ldots \}}$ for some $\man{\Phi}$. This follows from the definition of $\mathit{mtype}$ that searches \m{E}'s refinement chain from right to left if \m{m} is not declared in \m{C}'s refinement chain. Second, if \m{m} is defined in the declaration or in any refinement of class \m{C}, then $\mtypeY{m}{\mathit{last}(\masf{C})}$ should also be the same as $\mtypeY{m}{\mathit{last}(\masf{E})}$ with  $\mathit{CT}(\qtm{\Phi}{C}) = \masf{class C extends E \{ \ldots \}}$ for some $\man{\Phi}$. This case is covered by the well-formedness rules for methods that use the predicate $\mathit{override}$ to ensure that \m{m} is properly overridden, i.e., the signatures of the overridden and the overriding declaration of \m{m} are equal, and that \m{m} is not introduced twice, i.e., overloading is not allowed in \ffj. \qed
\end{proof}
\begin{lem}[\textit{Term substitution preserves typing}]
\label{lem:2}
If $\man{\Gamma}, \masf{\ol{x : B}} \vdash \masf{t : D}$ and $\man{\Gamma}, \masf{\ol{s : A}}$, where $\masf{\ol{A <: B}}$, then $\man{\Gamma} \vdash \left[\overline{\masf{x} \mapsto \masf{s}} \right] \masf{t : C}$ for some $\masf{C <: D}$.
\end{lem}
\begin{proof}
By induction on the derivation of $\man{\Gamma}, \masf{\ol{x : B}} \vdash \masf{t : D}$. 
\begin{case_}[\textsc{T-Var}]
$\ \masf{t} = \masf{x} \quad \masf{x : D} \in \man{\Gamma}$
\end{case_}
If $\masf{x} \not\in \mol{x}$, then the result is trivial since $\xss \masf{x} = \masf{x}$.\footnote{Note that $\xss \masf{x}$ is an abbreviation for $\left[\masf{x}_1 \mapsto \masf{s}_1, \ldots, \masf{x}_n \mapsto \masf{s}_n \right] \masf{x}$. It means that all occurrences of the variables $\masf{x}_1, \ldots, \masf{x}_n$ in the term $\masf{x}$ are substituted with the corresponsing terms $\masf{s}_1, \ldots, \masf{s}_n$.} On the other hand, if $\masf{x} = \masf{x}_i$ and $\masf{D} = \masf{B}_i$, then, since $\xss \masf{x} = \masf{s}_i$, letting $\masf{C} = \masf{A}_i$ finishes the case.
\begin{case_}[\textsc{T-Field}]
$\ \masf{t} = \masf{t$_0$.f$_i$} \quad \man{\Gamma}, \masf{\ol{x : B}} \vdash \masf{t$_0$ : D$_0$} \quad \fieldsY{\mathit{last}(\masf{D$_0$})} = \mol{C~f} \quad \masf{D} = \masf{C}_i$
\end{case_}
By the induction hypothesis, there is some $\masf{C}_0$ such that $\man{\Gamma} \vdash \left[\overline{\masf{x} \mapsto \masf{s}} \right] \masf{t$_0$ : C$_0$}$ and $\masf{C$_0$ <: D$_0$}$. It is easy to check that $\fieldsY{\mathit{last}(\masf{C$_0$})} = (\fieldsY{\mathit{last}(\masf{D$_0$}}), \mol{D~g})$ for some $\mol{D~g}$. Therefore, by \textsc{T-Field}, $\man{\Gamma} \vdash (\xss \masf{t$_0$})\masf{.f$_i$ : C$_i$}$. The fact that the refinements of a class may add new fields does not cause problems. $\mol{D~g}$ contains all fields that $\masf{C$_0$}$, including all of its refinements, add to $\masf{D$_0$}$.
\begin{case_}[\textsc{T-Invk}]
$\ \masf{t} = \masf{t$_0$.m(\ol{t})} \quad \man{\Gamma}, \masf{\ol{x : B}} \vdash \masf{t$_0$ : D$_0$} \quad \mtypeY{m}{\mathit{last}(\masf{D$_0$})} = \mol{E} \fct \masf{D} \\ \man{\Gamma}, \masf{\ol{x : B}} \vdash \masf{\ol{t : D}} \quad \masf{\ol{D <: E}}$
\end{case_}
By the induction hypothesis, there are some $\masf{C$_0$}$ and $\mol{C}$ such that:
\begin{displaymath}
\man{\Gamma} \vdash \left[\overline{\masf{x} \mapsto \masf{s}} \right] \masf{t}_0 \masf{ : C}_0 \quad \masf{C}_0 \masf{ <: D}_0 \quad \man{\Gamma} \vdash \xss \masf{\ol{t : C}} \quad \masf{\ol{C <: D}}.
\end{displaymath}
By Lemma~\ref{lem:1}, we have $\mtypeY{m}{\mathit{last}(\masf{C$_0$})} = \mol{E} \fct \masf{D}$. Moreover, $\masf{\ol{C <: E}}$ by the transitivity of $\masf{<:}\,$. Therefore, by \textsc{T-Invk}, $\man{\Gamma} \vdash \xss \masf{t$_0$.m($\xss$ \ol{t}) : D}$. The key is that subclasses and refinements may override methods but the well-formedness rules of methods ensure that the method's type is not altered, i.e., there is no overloading in \ffj.
\begin{case_}[\textsc{T-New}]
$\ \masf{t} = \masf{new D(\ol{t})} \quad \fieldsY{\mathit{last}(\masf{D})} = \mol{D~f} \quad \man{\Gamma}, \masf{\ol{x : B}} \vdash \masf{\ol{t : C}} \quad \masf{\ol{C <: D}}$
\end{case_}
By the induction hypothesis, $\man{\Gamma} \vdash \xss \masf{\ol{t : E}}$ for some $\mol{E}$ with $\masf{\ol{E <: C}}$. We have $\masf{\ol{E <: D}}$ by the transitivity of $\masf{<:}\,$. Therefore, by rule \textsc{T-New}, $\man{\Gamma} \vdash \masf{new D($\xss$ \ol{t}) : D}$. Although refinements of class \m{D} may add new fields, rule \textsc{T-New} ensures that the arguments of the object creation match the overall fields of \m{D}, including all refinements, in number and types. That is, the number of arguments ($\mol{t}$) equals the number of fields ($\mol{f}$) which function $\mathit{fields}$ returns.
\begin{case_}[\textsc{T-UCast}]
$\ \masf{t} = \masf{(D)t$_0$} \quad \man{\Gamma}, \masf{\ol{x : B}} \vdash \masf{t$_0$ : C} \quad \masf{C <: D}$
\end{case_}
By the induction hypothesis, there is some $\masf{E}$ such that $\man{\Gamma} \vdash \xss \masf{t$_0$ : E}$ and $\masf{E <: C}$. We have $\masf{E <: D}$ by the transitivity of $\masf{<:}\,$, which yields $\man{\Gamma} \vdash \masf{(D)(}\xss \masf{t$_0$) : D}$ by \textsc{T-UCast}. 
\begin{case_}[\textsc{T-DCast}]
$\ \masf{t} = \masf{(D)t$_0$} \quad \man{\Gamma}, \masf{\ol{x : B}} \vdash \masf{t$_0$ : C} \quad \masf{D <: C} \quad \masf{D} \neq \masf{C}$
\end{case_}
By the induction hypothesis, there is some $\masf{E}$ such that $\man{\Gamma} \vdash \xss \masf{t$_0$ : E}$ and $\masf{E <: C}$. If $\masf{E <: D}$ or $\masf{D <: E}$, then $\man{\Gamma} \vdash \masf{(D)($\xss$ t$_0$) : D}$ by \textsc{T-UCast} or \textsc{T-DCast}, respectively. If both $\masf{D $\not$<: E}$ and $\masf{E $\not$<: D}$, then $\man{\Gamma} \vdash \masf{(D)($\xss$ t$_0$) : D}$ (with a $\mathit{stupid\ warning}$) by \textsc{T-SCast}.
\begin{case_}[\textsc{T-SCast}]
$\ \masf{t} = \masf{(D)t$_0$} \quad \man{\Gamma}, \masf{\ol{x : B}} \vdash \masf{t$_0$ : C} \quad \masf{D $\not$<: C} \quad \masf{C $\not$<: D}$
\end{case_}
By the induction hypothesis, there is some $\masf{E}$ such that $\man{\Gamma} \vdash \xss \masf{t$_0$ : E}$ and $\masf{E <: C}$. This means that $\masf{E $\not$<: D}$ because, in \ffj, each class has just one superclass and, if both $\masf{E <: C}$ and $\masf{E <: D}$, then either $\masf{C <: D}$ or $\masf{D <: C}$, which contradicts the induction hypothesis. So $\man{\Gamma} \vdash \masf{(D)(}\xss \masf{t$_0$) : D}$ (with a $\mathit{stupid\ warning}$), by \textsc{T-SCast}. \qed
\end{proof}
\begin{lem}[\textit{Weakening}]
\label{lem:3}
If $\man{\Gamma} \vdash \masf{t : C}$, then $\man{\Gamma}, \masf{x : D} \vdash \masf{t : C}$
\end{lem}
\begin{proof}
Straightforward induction. The proof for \ffj is similar to the proof for \fj. \qed
\end{proof}
\begin{lem}
\label{lem:4}
If $\mtypeY{m}{\mathit{last}(\masf{C$_0$})} = \mol{D} \fct \masf{D}$, and $\mbodyB{m}{\mathit{last}(\masf{C$_0$})} = (\mol{x}, \masf{t})$, then for some $\masf{D}_0$ and some $\masf{C <: D}$ we have $\masf{C$_0$ <: D$_0$}$ and $\masf{\ol{x : D}}, \masf{this : D$_0$} \vdash \masf{t : C}\,$.
\end{lem}
\begin{proof}
By induction on the derivation of $\mbodyB{m}{\mathit{last}(\masf{C$_0$})}$. The base case (in which $\masf{m}$ is defined in the most specific refinement of $\masf{C$_0$}$) is easy since $\masf{m}$ is defined in $\mathit{CT}(\mathit{last}(\masf{C$_0$}))$ and the well-formedness of the class table implies that we must have derived $\masf{\ol{x : D}}, \masf{this : C$_0$} \vdash \masf{t : C}$ by the well-formedness rules of methods. The induction step is also straightforward: if \m{m} is not defined in $\mathit{CT}(\mathit{last}(\masf{C$_0$}))$, then $\mathit{mbody}$ searches the refinement chain from right to left; if \m{m} has not been found, the superclass' refinement chain is searched. There are two subcases: first, \m{m} is defined in the declaration or in any refinement of $\masf{C$_0$}$; this case is similar to the base case. Second, \m{m} is defined in a superclass $\masf{D$_0$}$ of $\masf{C$_0$}$ or in one of $\masf{D$_0$}$'s refinements; in this case, the well-formedness of the class table implies that we must have derived $\masf{\ol{x : D}}, \masf{this : D$_0$} \vdash \masf{t : C}$ by the well-formedness rules of methods, which finishes the case. \qed
\end{proof}

Note that this lemma holds because method refinements do not change the types of the arguments and the result of a method, overloading is not allowed, and $\masf{this}$ points always to the class that is introduced or refined.

\begin{thm}[\textit{Preservation}]
\label{thm:1}
If $\man{\Gamma} \vdash \masf{t : C}$ and $\masf{t $\longrightarrow$ t}'$, then $\man{\Gamma} \vdash \masf{t}' \masf{: C}'$ for some $\masf{C}' \masf{<: C}$.
\end{thm}

\begin{proof}
By induction on a derivation of $\masf{t} \longrightarrow \masf{t}'$, with a case analysis on the final rule.
\begin{case_}[\textsc{E-ProjNew}]
$\ \masf{t} = \masf{new C$_0$(\ol{v}).f$_i$} \quad \masf{t}' = \masf{v$_i$} \quad \fieldsY{\mathit{last}(\masf{C$_0$})} = \mol{D~f}$
\end{case_}
From the shape of $\masf{t}$, we see that the final rule in the derivation of $\man{\Gamma} \vdash \masf{t : C}$ must be \textsc{T-Field}, with premise $\man{\Gamma} \vdash \masf{new C$_0$(\ol{v}) : \masf{D$_0$}}$, for some $\masf{D$_0$}$, and that $\masf{C} = \masf{D$_i$}$. Similarly, the last rule in the derivation of $\man{\Gamma} \vdash \masf{new C$_0$(\ol{v}) : \masf{D$_0$}}$ must be \textsc{T-New}, with premises $\man{\Gamma} \vdash \masf{\ol{v : C}}$ and $\masf{\ol{C <: D}}$, and with $\masf{D$_0$} = \masf{C$_0$}$. In particular, $\man{\Gamma} \vdash \masf{v$_i$ : C$_i$}$, which finishes the case, since $\masf{C$_i$ <: D$_i$}$.
\begin{case_}[\textsc{E-InvkNew}]
$\ \masf{t} = \masf{(new C$_0$(\ol{v})).m(\ol{u})} \quad \masf{t}' = \left[ \overline{\masf{x} \mapsto \masf{u}}, \masf{this} \mapsto \masf{new C$_0$(\ol{v})} \right] \masf{t$_0$} \\ \mbodyB{m}{\mathit{last}(\masf{C$_0$})} = (\mol{x}, \masf{t$_0$})$
\end{case_}
The final rules in the derivation of $\man{\Gamma} \vdash \masf{t : C}$ must be \textsc{T-Invk} and \textsc{T-New}, with premises $\man{\Gamma} \vdash \masf{new C$_0$(\ol{v}) : C$_0$}$, $\man{\Gamma} \vdash \masf{\ol{u : C}}$, $\masf{\ol{C <: D}}$, and $\mtypeY{m}{\mathit{last}(\masf{C$_0$})} = \mol{D} \fct \masf{C}$. By Lemma~\ref{lem:4}, we have $\masf{\ol{x : D}}, \masf{this : D$_0$} \vdash \masf{t : B}$ for some $\masf{D$_0$}$ and $\masf{B}$, with $\masf{C$_0$ <: D$_0$}$ and $\masf{B <: C}$. By Lemma~\ref{lem:3}, $\man{\Gamma}, \masf{\ol{x : D}}, \masf{this : D$_0$} \vdash \masf{t$_0$ : B}$. Then, by Lemma~\ref{lem:2}, we have $\man{\Gamma} \left[ \overline{\masf{x} \mapsto \masf{u}}, \masf{this} \mapsto \masf{new C$_0$(\ol{v})} \right] \masf{t$_0$ : E}$ for some $\masf{E <: B}$. By the transitivity of $\masf{<:}\,$, we obtain $\masf{E <: C}$. Letting $\masf{C}' = \masf{E}$ completes the case.
\begin{case_}[\textsc{E-CastNew}]
$\ \masf{t} = \masf{(D)(new C$_0$(\ol{v}))} \quad \masf{C$_0$ <: D} \quad \masf{t}' = \masf{new C$_0$(\ol{v})}$
\end{case_}
The proof of $\man{\Gamma} \vdash \masf{(D)(new C$_0$(\ol{v})) : C}$ must end with \textsc{T-UCast} since ending with \textsc{T-SCast} or \textsc{T-DCast} would contradict the assumption of $\masf{C$_0$ <: D}$. The premises of \textsc{T-UCast} give us $\man{\Gamma} \vdash \masf{new C$_0$(\ol{v}) : C$_0$}$ and $\masf{D} = \masf{C}$, finishing the case.

The cases for the congruence rules are easy. We show just the case \textsc{E-Cast}.
\begin{case_}[\textsc{E-Cast}]
$\masf{t} = \masf{(D)t$_0$} \quad \masf{t}' = \masf{(D)t$'_0$} \quad \masf{t$_0$} \longrightarrow \masf{t$'_0$}$
\end{case_}
There are three subcases according to the last typing rule used.

\begin{subcase}[\textsc{T-UCast}]
$\ \man{\Gamma} \vdash \masf{t$_0$ : C$_0$} \quad \masf{C$_0$ <: D} \quad \masf{D} = \masf{C}$
\end{subcase}
By the induction hypothesis, $\man{\Gamma} \vdash \masf{t$'_0$ : C$'_0$}$ for some $\masf{C$'_0$ <: C$_0$}$. By transitivity of $\masf{<:}\,$, $\masf{C$'_0$ <: C}$. Therefore, by \textsc{T-UCast}, $\man{\Gamma} \vdash \masf{(C)t$'_0$ : C}$ (with no additional $\mathit{stupid\ warning}$).

\begin{subcase}[\textsc{T-DCast}]
$\ \man{\Gamma} \vdash \masf{t$_0$ : C$_0$} \quad \masf{D <: C$_0$} \quad \masf{D} = \masf{C}$
\end{subcase}
By the induction hypothesis, $\man{\Gamma} \vdash \masf{t$'_0$ : C$'_0$}$ for some $\masf{C$'_0$ <: C$_0$}$. If $\masf{C$'_0$ <: C}$ or $\masf{C <: C$'_0$}$, then $\man{\Gamma} \vdash \masf{(C)t$'_0$ : C}$ by \textsc{T-UCast} or \textsc{T-DCast} (without any additional $\mathit{stupid\ warning}$). On the other hand, if both $\masf{C$'_0$ $\not$<: C}$ or $\masf{C $\not$<: C$'_0$}$, then $\man{\Gamma} \vdash \masf{(C)t$'_0$ : C}$ with a $\mathit{stupid\ warning}$ by \textsc{T-SCast}.

\begin{subcase}[\textsc{T-SCast}]
$\ \man{\Gamma} \vdash \masf{t$_0$ : C$_0$} \quad \masf{D $\not$<: C$_0$} \quad \masf{C$_0$ $\not$<: D} \quad \masf{D} = \masf{C}$ 
\end{subcase}
By the induction hypothesis, $\man{\Gamma} \vdash \masf{t$'_0$ : C$'_0$}$ for some $\masf{C$'_0$ <: C$_0$}$. Then, also $\masf{C$'_0$ $\not$<: C}$ and $\masf{C $\not$<: C$'_0$}$. Therefore $\man{\Gamma} \vdash \masf{(C)t$'_0$ : C}$ with a $\mathit{stupid\ warning}$. If $\masf{C$'_0$ $\not$<: C}$, then $\masf{C $\not$<: C$'_0$}$ since $\masf{C $\not$<: C$_0$}$ and, therefore, $\man{\Gamma} \vdash \masf{(C)t$'_0$ : C}$ with $\mathit{stupid}\ \mathit{war\-ning}$. If $\masf{C$'_0$ <: C}$, then $\man{\Gamma} \vdash \masf{(C)t$'_0$ : C}$  by \textsc{T-UCast} (with no additional $\mathit{stupid\ warning}$). This subcase is analogous to the case \textsc{T-SCast} of the proof of Lemma~\ref{lem:2}. \qed
\end{proof}

\begin{thm}[\textit{Progress}]
\label{thm:2}
Suppose $\masf{t}$ is a well-typed term. 
\begin{enumerate}
\item If $\masf{t}$ includes $\masf{new C$_0$(\ol{t}).f$_i$}$ as a subterm, then $\fieldsY{\mathit{last}(\masf{C$_0$})} = \mol{C~f}$ for some $\mol{C}$ and $\mol{f}$.
\item If $\masf{t}$ includes $\masf{new C$_0$(\ol{t}).m(\ol{u})}$ as a subterm, then $\mbodyB{m}{\mathit{last}(\masf{C$_0$})} = (\mol{x}, \masf{t$_0$})$ and $\vert \mol{x} \vert = \vert \mol{u} \vert$ for some $\mol{x}$ and $\masf{t$_0$}$.
\end{enumerate}
\end{thm}

\begin{proof}
If $\masf{t}$ has $\masf{new C$_0$(\ol{t}).f$_i$}$ as a subterm, then, by well-typedness of the subterm, it is easy to check that $\fieldsY{\mathit{last}(\masf{C$_0$})}$ is well-defined and $\masf{f$_i$}$ appears in it. The fact that refinements may add fields (that have not been defined already) does not invalidate this conclusion. Note that for every field of a class, including its superclasses and all its refinements, there must be a proper argument. Similarly, if $\masf{t}$ has $\masf{new C$_0$(\ol{t}).m(\ol{u})}$ as a subterm, then it is also easy to show that $\mbodyB{m}{\mathit{last}(\masf{C$_0$})} = (\mol{x}, \masf{t$_0$})$ and $\vert \mol{x} \vert = \vert \mol{u} \vert$ from the fact that $\mtypeY{m}{\mathit{last}(\masf{C$_0$})} = \mol{C} \fct \masf{D}$ where $\vert \mol{x} \vert = \vert \mol{C} \vert$. This conclusion holds for \ffj since a method refinement must have the same signature than the method refined and overloading is not allowed. \qed
\end{proof}

\begin{thm}[\textit{Type soundness of} \ffj]
\label{thm:3}
If $\varnothing \vdash \masf{t : C}$ and $\masf{t} \longrightarrow^\ast \masf{t}'$ with $\masf{t}'$ a normal form, then $\masf{t}'$ is either a value $\masf{v}$ with  $\varnothing \vdash \masf{v : D}$ and $\masf{D <: C}$, or a term containing $\masf{(D)(new C(\ol{t}))}$ in which $\masf{C <: D}$.
\end{thm}

\begin{proof}
Immediate from Theorem~\ref{thm:1} and~\ref{thm:2}. Nothing changes in the proof of Theorem~\ref{thm:3} for \ffj compared to \fj. \qed
\end{proof}

\section{Type Soundness Proof of \ffjpl}
\label{app:ffjpl_ts}

In this section, we provide proof sketches of the theorems \textit{Correctness of} \ffjpl and \textit{Completeness of} \ffjpl. A further formalization would be desirable, but we have stopped at this point. As is often the case with formal systems, there is a trade-off between formal precision and legibility. We decided that a semi-formal development of the proof strategies are the best fit for our purposes.

\subsection{Correctness}
\label{sec:corr}

\begin{thm}[\textit{Correctness of} \ffjpl]
\label{thm:ffjpl_correct_app}
\normalfont Given a well-typed \ffjpl product line $\mathit{pl}$ (including with a well-typed term $\masf{t}$, well-formed class, introduction, and refinement tables $\mathit{CT}$, $\mathit{IT}$, and $\mathit{RT}$, and a feature model $\mathit{FM}$), every program that can be derived with a valid feature selection $\mathit{fs}$ is a well-typed \ffj program (cf.~Figure~\ref{fig:derivation}).
\begin{equation*}
\dfrac{
\mathit{pl} = (\masf{t}, \mathit{CT},\mathit{IT},\mathit{RT},\mathit{FM}) \qquad \mathit{pl} \mbox{\normalfont~is well-typed } \qquad \mathit{fs} \mbox{\normalfont~is valid in } \mathit{FM}
}
{
\mathit{derive}(\mathit{pl},\mathit{fs}) \mbox{\normalfont~is well-typed}
}
\end{equation*}
\end{thm}

The proof strategy is as follows: assuming that the \ffjpl type system ensures that each slice is a valid \ffj type derivation (Lemma~\ref{lem:a}) and that each valid feature selection corresponds to a single slice (Lemma~\ref{lem:a}), it follows that the corresponding feature-oriented program is well-formed. Before we prove Theorem~\ref{thm:ffjpl_correct_app} we develop two required lemmas that cover the two assumptions of our proof strategy.

\begin{lem}
\label{lem:a}
Given a well-formed \ffjpl product line, every slice of the product line's type derivation corresponds to a (set of) valid type derivation(s) in \ffj.
\end{lem}

\begin{proof}[Proof sketch]
Given a well-formed \ffjpl product line, the corresponding type derivation consists of possibly multiple slices. 

The basic case is easy: there is only a simple derivation without branches due to mutually exclusive features (optional features may be present). In this case, each term has only a single type, which is the one that would also be determined by \ffj. Furthermore, \ffjpl guarantees that referenced types, methods, and fields are present in all valid variants, using the predicate $\mathit{\validref}$. 

Let us illustrate this with the rule \textsc{T-Field}$_{\mathit{PL}}$; the other rules are analogous: 
\begin{equation*}
\qquad \dfrac{
\begin{array}{c}
\forall\, \masf{E} \in \mol{E} \, : \, \mathit{\validref}(\man{\Phi},\masf{E.f}) \\
\man{\Gamma}\ \vdash\ \masf{t$_0$}\ \masf{: } \mol{E} \dashv \man{\Phi} \qquad \mathit{fields}(\man{\Phi},\overline{\mathit{last}(\masf{E})}) = \overline{\overline{\f,\masf{C~f},\g}}
\end{array}
}
{
\man{\Gamma}\ \vdash\ \masf{t$_0$} \masf{.f}\ \masf{: } \masf{C}_{11}, \ldots, \masf{C}_{n1}, \ldots, \masf{C}_{1m}, \ldots, \masf{C}_{nm}  \dashv \man{\Phi}
} \qquad (\mbox{\sc T-Field}_{\mathit{PL}})
\end{equation*}

In the basic case there are no branches in the type derivation and thus the term $\masf{t$_0$}$ has only a single type $\masf{E$_1$}$. For the same reason, $\mathit{fields}$ returns only a simple list of fields that contains the declaration of field $\masf{f}$. Finally, \textsc{T-Field}$_{\mathit{PL}}$ checks whether the declaration of $\masf{f}$ is present in all valid variants (using $\mathit{\validref}$). Hence, in the basic case, an \ffjpl derivation that ends at the rule \textsc{T-Field}$_{\mathit{PL}}$ is equivalent to a set of corresponding \ffj derivations, which do not contain alternative and optional features and thus $\masf{t$_0$}$ has a single type, $\mathit{fields}$ returns a simple list of fields that contains the declaration of $\masf{f}$, and the declaration of $\masf{f}$ is present. The reason that an \ffjpl derivation without mutually exclusive features (i.e., a single slice) corresponds to multiple \ffj derivations is that the \ffjpl derivation may contain optional features whose different combinations correspond to the different \ffj derivations. Using predicate $\mathit{\validref}$, all type rules of \ffjpl ensure that all possible combinations of optional features are well-typed. 

In the case that there are multiple slices in the \ffjpl derivation, a term $\masf{t$_0$}$ may have multiple types $\mol{E}$. The type rules of \ffjpl make sure that every possible shape of a given term is well-typed. Each possible type of the term leads to a branch in the derivation tree. The premise of \textsc{T-Field}$_{\mathit{PL}}$ checks whether all possible shapes of a given term are well-typed by taking the conjunction of all branches of the derivation. Hence, if \textsc{T-Field}$_{\mathit{PL}}$ is successful, each individual branch holds, i.e., each slice corresponds to a well-typed \ffj program. Ensuring that, in the presence of optional features, all relevant subterms are well-typed (i.e., all referenced elements are present in all valid variants), a well-typed slice covers a set of well-typed \ffj derivations that correspond to different combinations of optional features, like in the basic case.

For example, in a field projection $\masf{t$_0$.f}$, the subterm $\masf{t$_0$}$ has multiple types $\mol{E}$. For all these types, $\mathit{fields}$ yields all possible combinations of fields declared by the variants of the types. It is checked whether, for each type of the subterm $\masf{t$_0$}$, each combination of fields contains a proper declaration of field $\masf{f}$. The different types of $\masf{f}$ become the possible types of the overall field projection term. Like in the basic case, it is checked whether every possible type of $\masf{t$_0$}$ is present in all valid variants (using $\mathit{\validref}$), so that each slice corresponds a valid \ffj derivation, i.e., a whole set of derivations covering different combinations of optional features. \qed
\end{proof}

\begin{lem}
\label{lem:b}
Given a well-formed \ffjpl product line, each valid feature selection corresponds to a single slice in the corresponding type derivation.
\end{lem}

\begin{proof}[Proof sketch]
By definition, a valid feature selection does not contain mutually exclusive features. Considering only a single valid feature selection, each term has only a single type. But the type derivation of the overall product line contains branches corresponding to alternative types of the terms. A successive removal of mutually exclusive features removes these branches until only a single branch remains. Consequently, a valid feature selection corresponds to a single slice. \qed
\end{proof}

\begin{proof}[Proof sketch of Theorem~\ref{thm:ffjpl_correct_app} (\textit{Correctness of} \ffjpl)]
The fact that the \ffjpl type system ensures that each slice is a valid \ffj type derivation (Lemma~\ref{lem:a}) and that each valid feature selection corresponds to a single slice (Lemma~\ref{lem:b}), implies that each feature-oriented program that corresponds to a valid feature selection is well-formed. \qed
\end{proof}

\subsection{Completeness}

\begin{thm}[\textit{Completeness of} \ffjpl]
\label{thm:ffjpl_complete_app}
Given an \ffjpl product line $\mathit{pl}$ (including a well-typed term $\masf{t}$, well-formed class, introduction, and refinement tables $\mathit{CT}$, $\mathit{IT}$, and $\mathit{RT}$, and a feature model $\mathit{FM}$), and given that \emph{all} valid feature selections $\mathit{fs}$ yield well-typed \ffj programs, according to Theorem~\ref{thm:ffjpl_correct_app}, $\mathit{pl}$ is a well-typed product line according to the rules of \ffjpl.
\begin{equation*}
\dfrac{
\mathit{pl} = (\masf{t},\mathit{CT},\mathit{IT},\mathit{RT},\mathit{FM}) \quad \forall\,\mathit{fs}\,:\,( \mathit{fs} \mbox{\normalfont~is valid in } \mathit{FM} \Rightarrow \mathit{derive}(\mathit{pl},\mathit{fs}) \mbox{\normalfont~is well-typed} )
}
{
\mathit{pl} \mbox{\normalfont~is well-typed }
}
\end{equation*}
\end{thm}

\begin{proof}[Proof sketch of Theorem~\ref{thm:ffjpl_complete_app} (\textit{Completeness of} \ffjpl)]
There are three basic cases: (1) $\mathit{pl}$ has only mandatory features; (2) $\mathit{pl}$ has only mandatory features except a single optional feature; (3) $\mathit{pl}$ has only mandatory features except two mutually exclusive features. Proving Theorem~\ref{thm:ffjpl_complete_app} for the first basic case is trivial. Since only mandatory features exist, only a single \ffj program can be derived from the product line. If the \ffj program is well-typed, the product line is well-typed, too, because all elements are always reachable and each term has only a single type. In fact, the type rules of \ffjpl and \ffj become equivalent in this case.

In the second basic case, two \ffj programs can be derived from the product line, one including and one excluding the optional feature. The difference between the two programs is the content of the optional feature. The feature can add new classes, refine existing classes by new methods and fields, and refine existing methods by overriding. If the two programs are well-typed, then the overall product line is well-typed as well since the reachability checks succeed in every type rule of \ffjpl. Otherwise, at least one of the two programs would not be well-typed since, in this case, the reachability checks are the only difference between \ffjpl's and \ffj's type rules (as in the first case, each term has only a single type since there are no mutually exclusive features). The fact that the two \fj programs are well-typed implies that all elements are reachable in the type derivations of two \ffj programs. Thus, the reachability checks of the \ffjpl derivation succeed in every case, i.e., the product line in question is well-typed.

In the third basic case, two \ffj programs can be derived from the product line, one including the first alternative and the other including the second alternative of the feature in question. The difference between the two programs is, on the one hand, the program elements one feature introduces that are not present in the other and, on the other hand, the alternative definitions of similar elements, like two alternative definitions of a single class. The first kind of difference is already covered by the second basic case. Alternative definitions of a program element (second kind of difference) that are well-typed in the context of their enclosing \ffj programs, are well-typed in \ffjpl because they lead to two new branches in the derivation tree which are handled separately and the conjunction of their premises must hold. Since the corresponding \ffj type rule for the element succeeds in both \ffj programs, their conjunction in the \ffjpl type rule always holds, i.e., the product line in question is well-typed.

Finally, we it remains to show that all other cases, i.e., all other combinations of mandatory, optional, and alternative features, can be reduced to combinations of the three basic cases, which proves Theorem~\ref{thm:ffjpl_complete_app}. To this end, we divide the possible relations between features into three disjoint sets: (1) a feature is reachable from another feature in \emph{all} variants, (2) a feature is reachable from another feature in \emph{some}, but \emph{not in all}, variants, (3) two features are mutually exclusive. From these three possible relations we construct a general case that can be reduced to a combination of the three basic cases. 

Assume a feature $\man{\Phi}$ that is mandatory with respect to a set of features $\overline{\man{\Pi}}$, that is optional with respect to a set of features $\overline{\man{\Omega}}$, and that is alternative to a set $\overline{\man{\Delta}}$ of features. We use arrows to illustrate to which of the three basic cases a pairwise relation between $\man{\Phi}$ and each element of a list is reduced:
\begin{equation*}
\xymatrixcolsep{6ex}
\begin{xy}
  \xymatrix{
      \man{\Phi} \ar[r]^2 \ar[d]_1 \ar[dr]^3   &   \overline{\man{\Omega}} \\
      \overline{\man{\Pi}}                     &   \overline{\man{\Delta}}   
  }
\end{xy}
\end{equation*}
Such an arrow diagram can be created for every feature of a product line. The reason is that the three kinds of relations are orthogonal and  there are no further relations relevant for type checking. Hence, the general case covers all possible relations between features and combinations of features. The description of the general case and the reduction finish the proof of Theorem~\ref{thm:ffjpl_complete_app}, i.e., \ffjpl's type system is complete. \qed
\end{proof}

\end{appendix}

\end{document}